\documentclass{aastex7}

\newcommand{\rrab}{RR{\sl ab}}
\newcommand{\rrc}{RR{\sl c}}
\newcommand{\rrd}{RR{\sl d}}
\defcitealias{vivas20a}{V20}

\shorttitle{Is Crater II disrupting?}
\shortauthors{Vivas et al.}

\begin{document}

\title{Is Crater II disrupting?}

\correspondingauthor{A. K. Vivas}

\author[orcid=0000-0003-4341-6172,sname='Vivas']{A.~Katherina~Vivas}
\affiliation{Cerro Tololo Inter-American Observatory/NSF NOIRLab, Casilla 603, La Serena, Chile}
\email{kathy.vivas@noirlab.edu}

\author[0000-0002-7123-8943]{Alistair~R.~Walker}
\affiliation{Cerro Tololo Inter-American Observatory/NSF NOIRLab, Casilla 603, La Serena, Chile}
\email{alistair.walker@noirlab.edu}

\author[0000-0002-9144-7726]{Clara~E.~Mart{\'\i}nez-V\'azquez}
\affiliation{NSF NOIRLab, 670 N. A’ohoku Place, Hilo, Hawai’i, 96720, USA}
\email{clara.martinez@noirlab.edu}

\author[0009-0001-8169-5398]{MJ Cooke}
\affiliation{Department of Astronomy, The University of Texas at Austin, Austin, TX 78712, USA}
\email{mjc9700@gmail.com}

\author[0000-0001-6728-806X]{Carme Gallart}
\affiliation{IAC- Instituto de Astrof\'isica de Canarias, Calle V\'ia Lactea s/n, E-38205 La Laguna, Tenerife, Spain}
\affiliation{Departmento de Astrof\'isica, Universidad de La Laguna, E-38206 La Laguna, Tenerife, Spain}
\email{carme.gallart@iac.es}

\author[0000-0001-5292-6380]{Matteo Monelli}
\affiliation{INAF-Osservatorio Astronomico di Roma, via Frascati 33, 00040 Monte Porzio Catone, Italy}
\affiliation{IAC- Instituto de Astrof\'isica de Canarias, Calle V\'ia Lactea s/n, E-38205 La Laguna, Tenerife, Spain}
\affiliation{Departmento de Astrof\'isica, Universidad de La Laguna, E-38206 La Laguna, Tenerife, Spain}
\email{matteo.monelli@inaf.it}

\author[0009-0009-5159-9598]{Jaime A. Rojas Cancino}
\affiliation{Instituto de Astrof{\'\i}sica, Pontificia Universidad Cat\'olica de Chile, Av. Vicu\~na Mackenna 4860, 7820436 Macul, Santiago, Chile}
\email{jaimealfrojas@gmail.com}

\author[0000-0002-1793-3689]{David~L.~Nidever}
\affiliation{Department of Physics, Montana State University, P.O. Box 173840, Bozeman, MT 59717-3840, USA} 
\email{dnidever@montana.edu}

\begin{abstract}

Crater~II (CraII) is a very intriguing, low-surface brightness and extended galaxy in the vicinity of our Milky Way. Motivated by its huge area and the number of RR Lyrae stars (RRLS) detected near to its core, we performed a follow-up campaign to get deep multi-band ($g,i$) and multi-epoch DECam observations in the outskirts of CraII, covering up to 8$r_h$ in the North-East direction and up to 13$r_h$ in the South-West direction (spanning almost $10\degr$ in the sky across the galaxy). We confirm the existence of tails coming out from CraII. In our survey, we detected a total of 97 periodic variable stars, of which 46 are CraII members [37 RRLS and 7 anomalous Cepheids (AC)]. The RRLS show a strong gradient in distance of 3.7~kpc/deg, which follows the relation: $d (\rm {kpc}) = (-3.70\pm 0.21)\, \xi' + (116.2 \pm 0.32)$ (where $\xi'$ is the planar coordinate rotated to fit the CraII tails). We also found an indication that the gradient in distance in the distant part of the near-side tail (North-East) may be less steep than it is in the inner $5\deg$. The leading tail also displays a significant over-density of RRLS located $\sim 3.25$ deg from the center of CraII.
Despite covering a smaller area than previous works, our deeper photometry has allowed us to unveil more variable stars in the region and better define the tidal tails of CraII. 

\end{abstract}

\section{Introduction} \label{sec:intro}

Along with Antlia II \citep{torrealba19}, Crater II \citep[hereafter CraII,][]{torrealba16} exemplifies a class of ultra-diffuse, low-luminosity satellites that were previously unrecognized in the census of Milky Way companions. The properties of CraII challenge our understanding of galaxy formation and evolution. In particular, the galaxy exhibits an unusually low line-of-sight velocity dispersion of $\sigma_{\rm los} = 2.7$ km/s, as initially measured by \citet{caldwell17} and later confirmed by \citet{fu19}. This makes CraII one of the kinematically coldest stellar systems known, significantly below the typical dispersions ($\sim 10$ km/s) expected for galaxies of similar size \citep{caldwell17}.

Several theoretical models have been proposed to explain this discrepancy. While MOdified Newtonian Dynamics (MOND) could in principle account for the low dispersion \citep{mcgaugh16}, simulations within the $\Lambda$CDM framework suggest that substantial tidal stripping is required to suppress the velocity dispersion to such low levels. \citet{sanders18} argued that CraII may have lost about 70\% of its original mass, while \citet{fattahi18} proposed even more extreme mass loss scenarios, up to 99\%. With the advent of Gaia DR3, orbital modeling became possible, further supporting the case for strong tidal interactions. CraII is in an orbit around the Milky Way which has a pericentric distance of only 27-39 kpc \citep[depending on the assumed potential of the Milky Way,][]{battaglia22}, and a short-period ($\sim 1.8-3.0$ Gyr). Such repeated close encounters make significant tidal stripping not only plausible but likely \citep[see also][]{fritz18,fu19,pace22}.

Additional clues about CraII’s evolutionary history come from its stellar populations. Using deep Dark Energy Camera (DECam) photometry, \citet{walker19} constructed a color-magnitude diagram (CMD) revealing two distinct subgiant branches that converge into a narrow red giant branch. These features point to two separate episodes of star formation, approximately 10.5 and 12.5 Gyr ago, with minimal chemical enrichment between them. Spectroscopic studies \citep{caldwell17, fu19, Ji21} indicate that the galaxy is metal-poor, with an average metallicity of [Fe/H] $= –2.0$, yet surprisingly lacks a blue horizontal branch (BHB) population. This metallicity places CraII slightly below the luminosity-metallicity relationship seen in dwarf galaxies \citep{kirby13}, implying that CraII must have lost up to 42\% of its initial mass \citep{Ji21}.

There is also evidence that the younger stellar population is more centrally concentrated \citep{walker19}, potentially reflecting the complex interplay between internal evolution and external tidal forces. These observations raise compelling questions: Was an even older stellar population ($>12.5$ Gyr) preferentially stripped and is now undetectable? Was the formation of the younger population triggered by an early pericentric passage before the gas was fully removed? Could CraII be a recent accretion into the Milky Way halo?

To address these questions, it is essential to confirm whether CraII is currently undergoing tidal disruption and to constrain the extent of mass loss it may have experienced. Given the galaxy’s extremely low surface brightness and the large distance from us \citep[116.5 kpc,][hereafter V20]{vivas20a}, this task is observationally challenging. Recently, \citet{coppi24} used RRLS to trace the outskirts of the CraII galaxy finding indeed two long tails, extending $\sim 15\degr$ in the sky, with one tail being closer to us than the other. This discovery leaves no doubt that CraII is indeed disrupting.  

\citet{coppi24}'s work demonstrates, once again, that RRLS are an excellent tool to trace stellar populations in different stellar systems. In the past they have also been used to trace the tidal tails of the Sagittarius dSph galaxy \citep[eg.][]{vivas01,hernitschek17,muraveva25}, the outskirts of Antlia~II \citep{vivas22}, Carina \citep{vivas13} and Ursa Minor \citep{garofalo25}, the extended stellar populations in some Ultra-Faint Dwarf galaxies \citep{garling18,vivas2020b,tau24} and globular clusters \citep{price19,shipp20}, stellar streams \citep{duffau06,vivas16b,prudil21}, among others. RRLS are variable stars of an old age ($> 10$ Gyrs), and they are found in large numbers in CraII \citep[][\citetalias{vivas20a}]{joo18,monelli18}. They have the additional advantage of being excellent standard candles. Thus, any RRLS found in the neighborhood of CraII and having a similar distance as the galaxy are very likely associated with it. Although RRLS have been found in the galactic halo up to distances of $\sim 300$ kpc \citep{medina18,stringer21,medina24,feng24}, they are very rare at the large galactocentric distance of CraII. Thus, the contamination by Milky Way RRL field stars is expected to be minimal.

\citet{coppi24} discovered the CraII tails using the La Silla-QUEST RRLS survey \citep{zinn14}, a large variability survey made with the 1m Schmidt telescope of the La Silla Observatory. Their search around CraII covered an impressive area of 300 sq. deg. around the galaxy. The limiting magnitude of their survey, however, lies around the magnitude of the RRLS in CraII, and their completeness significantly drops after $\sim 120$ kpc.

In 2020 we initiated a project of time-series observations to look for RRLS in the outskirts of CraII using DECam at the 4m Blanco telescope at Cerro Tololo Inter-American Observatory, following our successful campaign to cover the central field of CraII \citepalias{vivas20a}. The combination of the 3 sq. deg. field of view of DECam with the large aperture of the Blanco telescope is an ideal combination to look for debris material around CraII. Once the results by \citet{coppi24} were known, we decided to extend the survey in the direction of the distant tail of CraII, where the completeness of the La Silla-QUEST survey may be particularly low. Our goal is to improve the characterization of the CraII tails with our deeper dataset by detecting a complete sample of RRLS and with them measure with precision the radial gradient of distance of the tails. In this paper we describe the observations in Section~\ref{sec:observations} and the search for periodic variable stars in Section~\ref{sec:variables}. In Section~\ref{sec:RRL} we discuss the spatial distribution of the RRLs around CraII and we characterize the distance gradient observed in the tails. Section~\ref{sec:AC} discuss the membership of Anomalous Cepheid (AC) stars along the tails. Finally, our discussion and conclusions are presented in Section~\ref{sec:conclusions}.

\section{Observations and Data Processing} \label{sec:observations}

\begin{deluxetable}{lcccccccccc}
\tablecolumns{11}
\tablewidth{0pc}
\tablecaption{Observed Fields \label{tab:fields}}
\tablehead{
\colhead{Field} & \colhead{RA (J2000.0)} & \colhead{DEC (J2000.0)} & \multicolumn{2}{c}{2020} & \multicolumn{2}{c}{2021} & \multicolumn{2}{c}{2024} & \multicolumn{2}{r}{Total epochs} \\
\colhead{} & \colhead{(deg)} & \colhead{(deg)} & \colhead{N$_g$} & \colhead{N$_i$} & \colhead{N$_g$} & \colhead{N$_i$} & \colhead{N$_g$} & \colhead{N$_i$} & \colhead{N$_g$} & \colhead{N$_i$} \\
}
\startdata
R1 & 179.06625 & -17.50083 & 19 & 26 & 12 & 13 &  &  & 31 & 39 \\
R2 & 177.52042 & -16.59389 & 19 & 27 & 11 & 12 &  &  & 30 & 39 \\
R3 & 175.84792 & -17.36389 & 18 & 25 & 11 & 12 &  &  & 29 & 37 \\
R4 & 175.52125 & -19.11333 & 18 & 25 & 11 & 11 &  &  & 29 & 36 \\
R5 & 176.94667 & -20.14472 & 18 & 25 & 11 & 11 &  &  & 29 & 36 \\
R6 & 178.63833 & -19.31361 & 18 & 25 & 11 & 12 &  &  & 29 & 37 \\
ER1 & 174.59917 & -20.53389 & 18 & 26 & 11 & 12 &  &  & 29 & 38 \\
ER2 & 180.29708 & -16.16472 & 19 & 26 & 12 & 13 &  &  & 31 & 39 \\
ER3 & 175.95458 & -21.63556 &  &  &  &  & 33 & 30 & 33 & 30 \\
ER4 & 173.38500 & -19.29694 &  &  &  &  & 33 & 29 & 33 & 29 \\
ER5 & 174.10917 & -22.48806 &  &  &  &  & 33 & 26 & 33 & 26 \\
ER6 & 172.68958 & -21.25111 &  &  &  &  & 30 & 26 & 30 & 26 \\
ER7 & 172.35292 & -23.04222 &  &  &  &  & 30 & 27 & 30 & 27 \\
\enddata
\end{deluxetable} 

\begin{figure}
\plotone{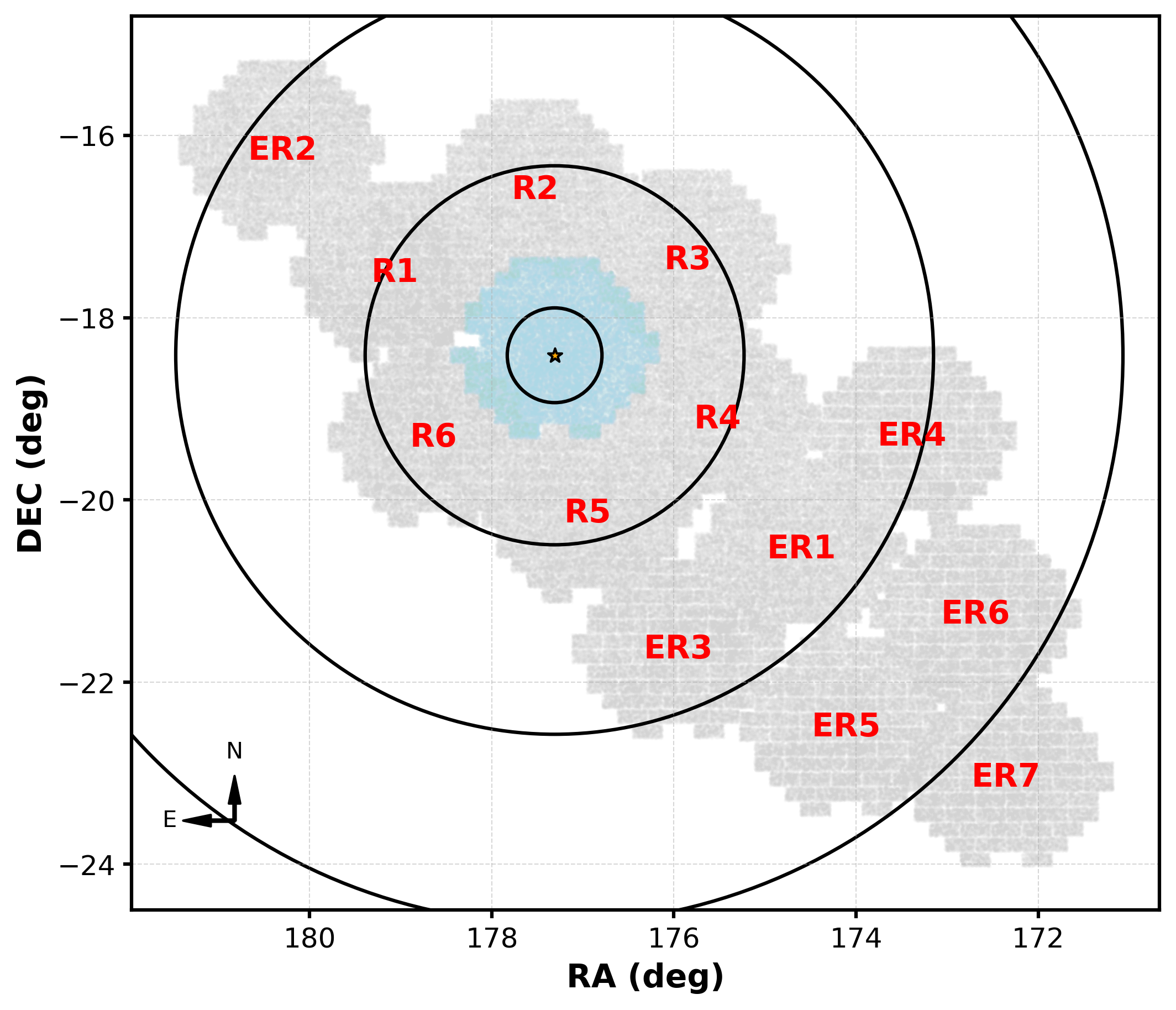}
\caption{Spatial coverage of our DECam observations of CraII in equatorial coordinates. The gray background, which shows the footprint of our survey, is a density map made with all stars in our catalog. The central region, in blue, shows the single field observed in \citetalias{vivas20a}. Black ellipses indicate $1\, r_h$, $4\, r_h$, $8\, r_h$, and $12\, r_h$. The orange star marks the center of CraII.}
\label{fig:fields}
\end{figure}

We used the DECam \citep{flaugher15} at the 4m Victor Blanco Telescope at Cerro Tololo Inter-American Observatory, Chile, to observe 13 external fields around CraII, as shown in Figure~\ref{fig:fields}. Fields R1 to R6 surround the galaxy, extending the uniform coverage of the galaxy to $\sim 4r_h$ \citep[$r_h=31\farcm 2$,][]{torrealba16}. Fields ER1 and ER2 were chosen to be located along the orbit of the galaxy. Finally fields ER3 to ER7 were added after the results by \citet{coppi24} appeared. These fields extend the coverage in the far-side tail (South-West direction) of CraII where the completeness of La Silla-QUEST survey drops significantly given that this side is farther from us ($>120$ kpc). The near-side tail is covered up to $\sim 8 r_h$, while the far-side tail is covered up to $\sim 13\, r_h$ from the center of the CraII.

Fields R1-R6 and ER1-ER2 were observed in 6 half-nights in 2020 (2020 Mar 4-7, Mar 15-16)\footnote{The 2020 campaign had to be interrupted due to the start of the COVID pandemic and the temporary closing of the Observatory.}, and 2 half-nights in 2021 (2021 Apr 29-30). For all these fields, we obtained consecutive $g$ and $i$ observations, alternating pointings with a $60\arcsec$ offset in each direction to cover the chip gaps. Fields ER3 to ER7 were observed during 5 partial nights in 2024 under Engineering and Director Discretionary Time (2024 Mar 27; Apr 1, 15; May 9-10). Given the limited amount of time available for these extra fields we did not dither to cover the chip gaps. In all cases, exposure times were 180s in each band, similar to what we used to observe the central field of CraII in \citetalias{vivas20a}. A total of 835 observations were obtained. Median seeing was $1\farcs 14$ and $0\farcs 98$ in $g$ and $i$, respectively. Table~\ref{tab:fields} contain the central coordinates of each field and the number of epochs obtained in each filter. On photometric nights, standard star fields in the SDSS footprint were observed at different airmasses. Thus, our photometry is tied to the SDSS system.

The data reduction, production of photometric catalogs, and selection of periodic variable stars follow exactly the same procedure used for our previous observations in the central field of CraII \citepalias{vivas20a}. We recommend the reader interested in these aspects to refer to Section~3 in \citetalias{vivas20a}. Raw and reduced data are publicly available in the NOIRLab Astro Data Archive\footnote{\url https://astroarchive.noirlab.edu} under propIDs 2020A-0058 and 2021A-0124.

At the level of the horizontal branch of CraII ($g\sim 21.0$; $i\sim 20.7$) our typical photometric errors are very low, 0.02 mag in both bands. Errors increase to 0.1~mag for magnitudes $g=23.2$, $i=23.0$. Thus, our data extends to $\sim 2$~mag below the horizontal branch, allowing us to discover any RRLS well beyond the distance of CraII.

 \section{Variable Stars} \label{sec:variables}

We searched for RRLS and AC stars following the variable star detection recipe described in detail in \citetalias{vivas20a}. We identified 97 of such periodic variable stars in the external fields of CraII, seven of which are located in the small overlapping region with the \citetalias{vivas20a}'s field (see Fig~\ref{fig:fields}). We kept the original ID number for those stars. We also recovered three stars (V24, V79 and V94) initially identified in \citet{joo18} that were outside the footprint of \citetalias{vivas20a}. All other stars were assigned ID numbers starting at V135 to continue with the sequence given in \citetalias{vivas20a}, with the numbering sequence increasing with the angular separation from the center of CraII. Table~\ref{tab:data} contains the assigned ID number, coordinates RA and DEC (J2000.0), period, type of variable, number of observations, amplitude, mean magnitude and mean error in the $g$ and $i$ bands, color excess E(B-V) from \citet{schlegel98}, angular separation from the center of CraII, distance modulus and its error (see Section~\ref{sec:gradient}), identification numbers for the stars in common with \citet{coppi24} and the Gaia DR3 catalog of RR Lyrae stars \citep{clementini23}, and comments on the light curves. Although Table~\ref{tab:data} contains all variables detected in this work, henceforth the seven stars in common are considered only as part of the central field stars in \citetalias{vivas20a} to avoid duplicates.    

\begin{figure}
\includegraphics[width=0.49\textwidth]{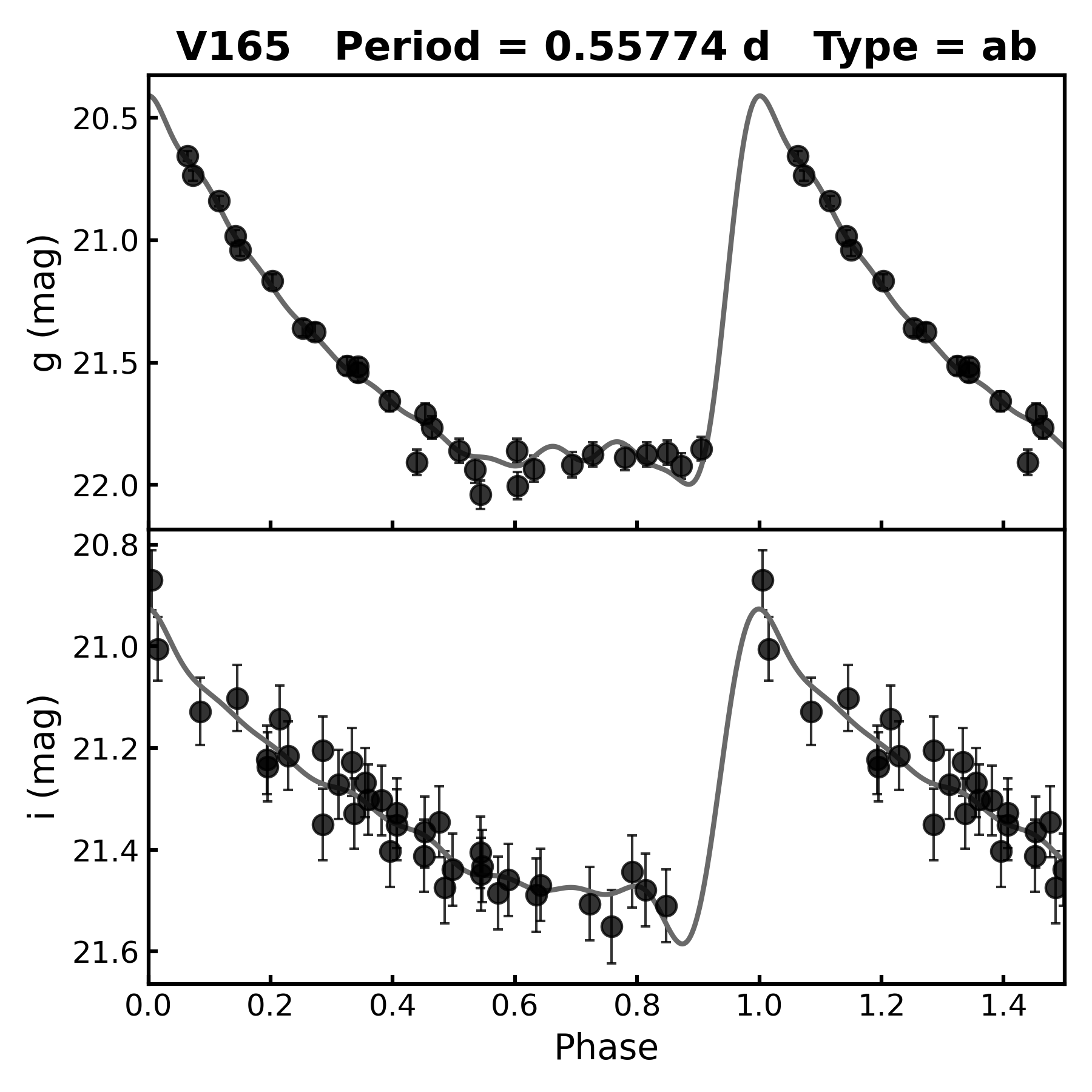}
\includegraphics[width=0.49\textwidth]{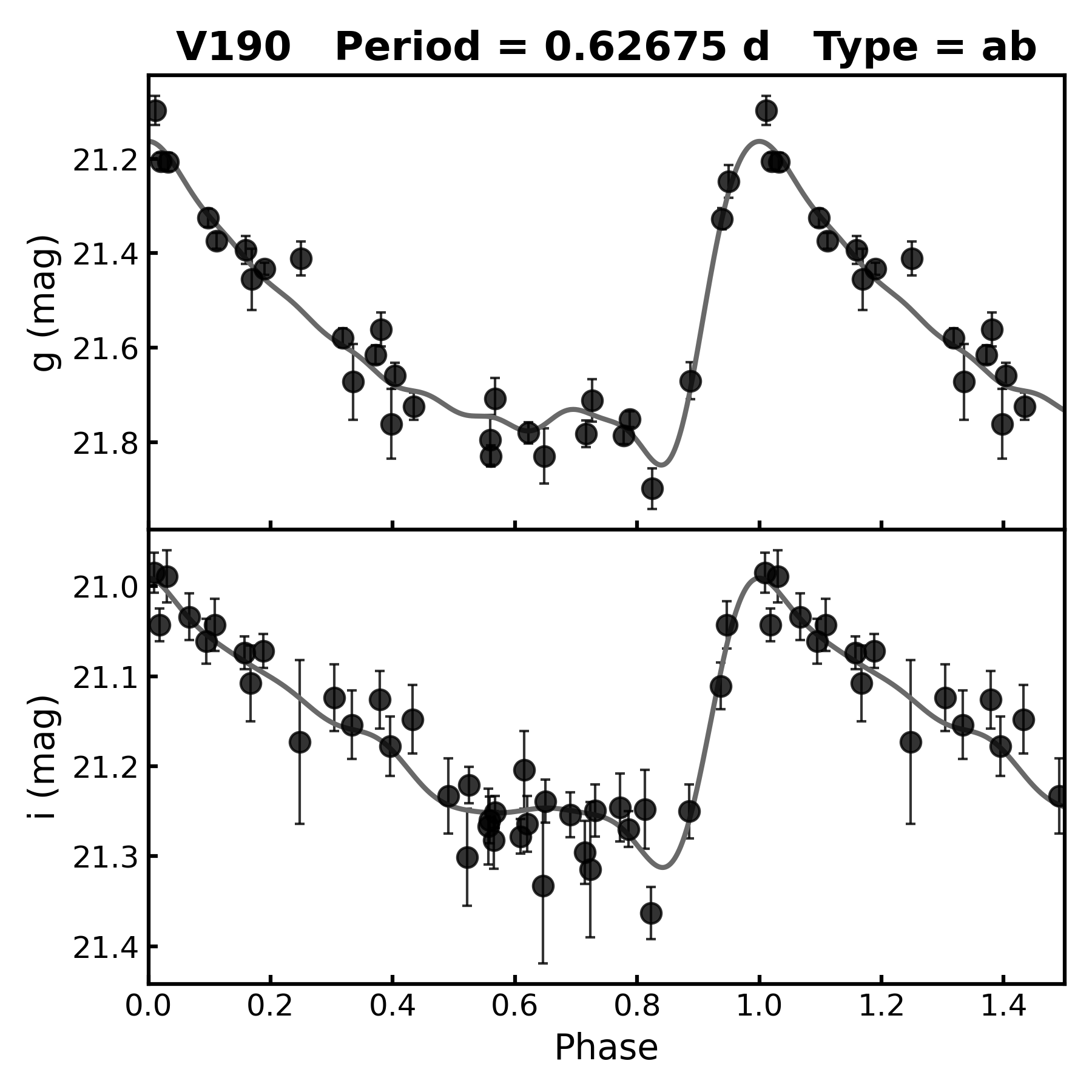}
\caption{Examples of the $g$, $i$ lightcurves of two of the faintest RRLS identified in this work. The solid lines are the best fitted template from the library of \citet{baeza25}. The full figure set of lightcurves is available in the online version of the journal.}
\label{fig:lc}
\end{figure}

Some examples of lightcurves are shown in Figure~\ref{fig:lc}. RRLS of the types ab and c were fitted with templates from the library of \citet{baeza25} using a custom-made wrapper of \texttt{gatspy} \citep{vanderplas15}. This library of templates is based on high-cadence observations of hundreds of RRLS in the DECam bands $griz$ and present a larger range of morphologies compared with the library of \citet{sesar10}, which is the one used in \citetalias{vivas20a}. Fitting templates is particularly useful for determining mean magnitudes when the number of observations is small and the sampling of the lightcurves is not uniform, both situations present in our data. Mean magnitudes for the {\rrab } and {\rrc} stars reported in Table~\ref{tab:data} were estimated by integrating the templates in intensity units and transforming the result back to magnitudes. We note that the mean magnitudes are not largely affected by the particular set of templates used. The mean difference in the mean $i$ magnitude when using the two sets of templates mentioned above is only $0.003 \pm 0.009$ mag. 

We did not fit templates to the {\rrd } stars nor to the AC stars. Mean magnitudes for those types are averages calculated in intensity units and transformed back to magnitudes. Table~\ref{tab:data} contains comments for noisy ligthcurves, poor template fits, or lightcurves missing observations near maximum or minimum light. In general the lightcurves and their template fits are very good and only a handful display such problems. Time-series photometry for all variable stars is provided in Table~\ref{tab:ts}.

Extinctions were calculated using $A_g = 3.303 \, \rm{E(B-V)}_{\rm SFD}$
and $A_i = 1.698 \, E(B - V)_{\rm SFD}$, where the coefficients were taken from \citet{schlafly11} so the extinctions are consistent with their recalibration of the \citeauthor{schlegel98}'s dust maps. All the explored region is low extinction, with values of E(B-V)$_{\rm SFD}$ ranging between 0.03 and 0.09~mag.

\begin{splitdeluxetable}{lccccccccccccBccccrrl}
\tabletypesize{\small}
\tablewidth{0pt} 
\tablecaption{Variable Stars in the Outskirts of CraII \label{tab:data}}
\tablehead{
\colhead{ID} & \colhead{RA (J2000.0)}& \colhead{DEC (J2000.0)} & \colhead{Period} &
\colhead{Type\tablenotemark{a}} & \colhead{N$_g$} & \colhead{Amp g} & \colhead{g} & \colhead{Error g} &
\colhead{N$_i$} & \colhead{Amp i} & \colhead{i} & \colhead{Error i} & \colhead{E(B-V)$_{\rm SFD}$} & \colhead{Angular Dist\tablenotemark{b}} & \colhead{$\mu_0$} & \colhead{error $\mu_0$} & \colhead{ID Coppi+24} & \colhead{ID Gaia DR2} & \colhead{Comments} \\
\colhead{} & \colhead{(deg)} & \colhead{(deg)} & \colhead{(d)} & \colhead{} 
& \colhead{} & \colhead{(mag)} & \colhead{(mag)} & \colhead{(mag)} & \colhead{} & 
\colhead{(mag)} & \colhead{(mag)} & \colhead{(mag)} &
\colhead{(mag)} & \colhead{(deg)} & \colhead{(mag)} & \colhead{(mag)} &
\colhead{} & \colhead{} & \colhead{} \\
} 
\startdata 
V91 & 177.867330 & -18.740810 & 0.41582 & c & 28 & 0.54 & 21.02 & 0.023 & 36 & 0.24 & 20.80 & 0.025 & 0.040 & 0.58 & 20.347 & 0.065 & & 3543912240464745856 &  \\
V54 & 178.071250 & -18.581360 & 0.58043 & ab & 28 & 1.26 & 20.98 & 0.025 & 36 & 0.61 & 20.75 & 0.026 & 0.035 & 0.70 & 20.319 & 0.065 & &  &  \\
V90 & 178.025320 & -18.735880 & 0.72261 & ab & 28 & 0.60 & 20.94 & 0.020 & 36 & 0.24 & 20.69 & 0.023 & 0.038 & 0.71 & 20.350 & 0.065 & &  & noisy \\
V115 & 177.835260 & -19.014320 & 0.50554 & Fab & 28 & 1.37 & 17.60 & 0.003 & 36 & 0.75 & 17.39 & 0.004 & 0.040 & 0.75 & 16.833 & 0.062 & & 3543062494071480448 &  \\
V24 & 177.176870 & -19.171220 & 0.62162 & ab & 23 & 0.83 & 21.13 & 0.025 & 29 & 0.42 & 20.78 & 0.025 & 0.047 & 0.77 & 20.362 & 0.065 & &  &  \\
V84 & 176.679330 & -17.814000 & 0.61525 & ab & 28 & 0.87 & 21.03 & 0.024 & 36 & 0.40 & 20.73 & 0.024 & 0.035 & 0.88 & 20.323 & 0.065 & &  &  \\
V37 & 178.268640 & -17.925360 & 0.59348 & ab & 29 & 1.00 & 20.96 & 0.024 & 39 & 0.45 & 20.71 & 0.024 & 0.033 & 0.99 & 20.295 & 0.065 & &  &  \\
V135 & 178.204530 & -17.744040 & 0.40956 & Fc & 28 & 0.55 & 16.26 & 0.003 & 24 & 0.25 & 16.11 & 0.003 & 0.034 & 1.05 & 15.595 & 0.061 & & 3568052331786266880 &  \\
\enddata
\tablenotetext{a}{Types: ab $=$ type ab RRLS; c $=$ type c RRLS; d $=$ type d RRLS; AC $=$ anomalous Cepheids; Fab $=$ type ab field RRLS; Fc $=$ type c field RRLS. Question mark (?) denotes uncertainty in the classification (see text)}
\tablenotemark{b}{Angular separation from the center of CraII, in degrees}
\tablecomments{Table~\ref{tab:data} is published in its entirety in the machine-readable format. A portion is shown here for guidance regarding its form and content.}
\end{splitdeluxetable}

\begin{deluxetable}{lcccc}
\tablecolumns{5}
\tablewidth{0pc}
\tablecaption{Time series photometry of variable stars in the outskirts of Crater II \label{tab:ts}}
\tablehead{
\colhead{ID} & \colhead{Filter} & \colhead{MJD} & \colhead{Mag} & \colhead{$\sigma$ Mag} \\
\colhead{} & \colhead{} & \colhead{(d)} & \colhead{(mag)} & \colhead{(mag)} \\
}
\startdata
 V91 & g  &  2458913.72969 & 20.885 &  0.021 \\
 V91 & g  &  2458913.88086 & 21.267 &  0.018 \\
 V91 & g  &  2458914.76486 & 21.291 &  0.033 \\
 V91 & g  &  2458914.86377 & 20.809 &  0.012 \\
 V91 & g  &  2458915.86501 & 21.012 &  0.013 \\
 V91 & g  &  2458916.78137 & 21.224 &  0.047 \\
\enddata
\tablecomments{Table~\ref{tab:ts} is published in its entirety in the machine-readable format. A portion is shown here for guidance regarding its form and content.}
\end{deluxetable} 

\section{RR Lyrae Stars in the Outskirts of Crater II} \label{sec:RRL}

\begin{figure}
\plotone{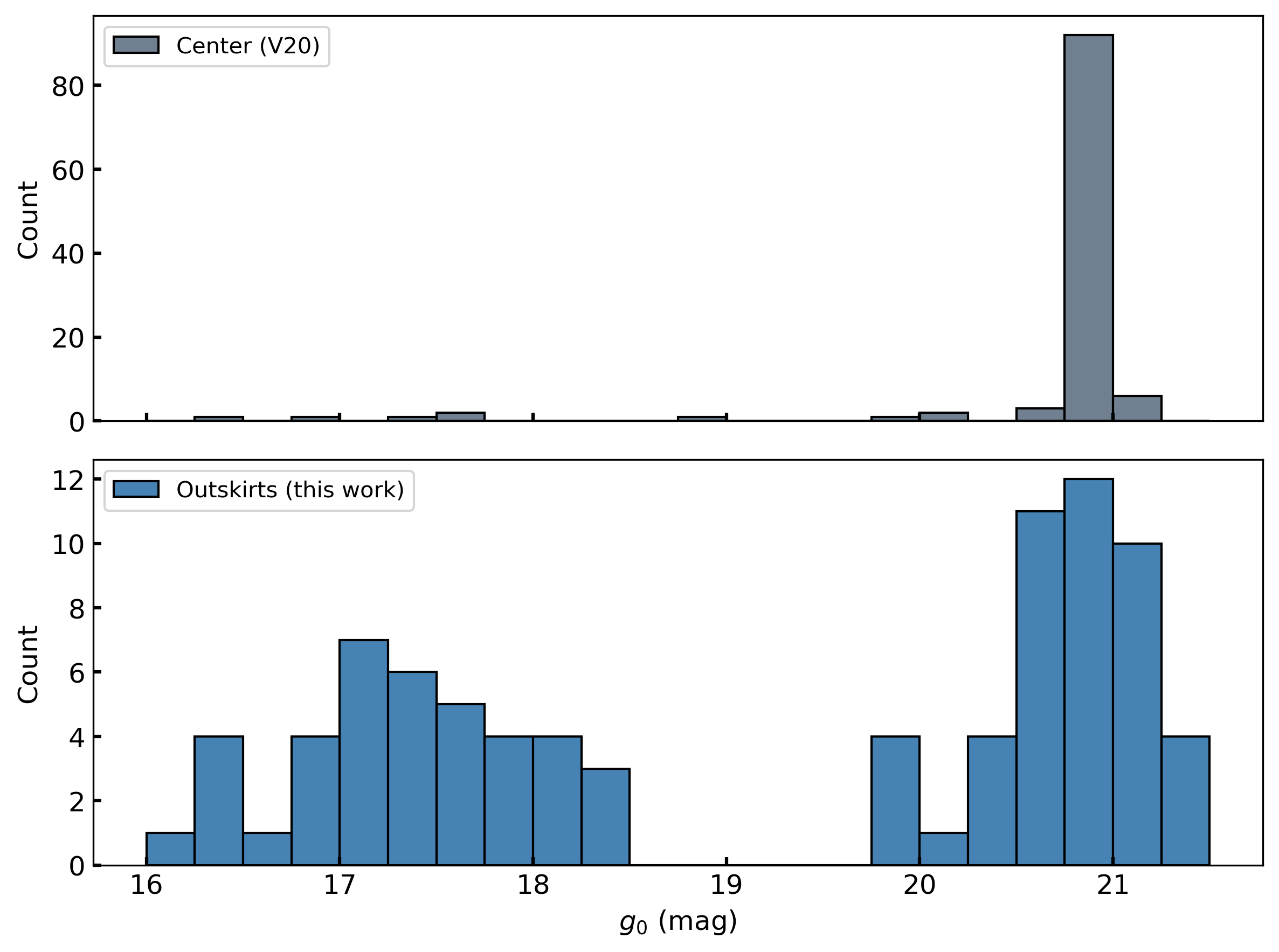}
\caption{Distribution of extinction-corrected $g$ magnitudes in the central field of CraII \citepalias[from][top histogram]{vivas20a}, and in the outskirts (bottom histogram).}
\label{fig:histo_g0}
\end{figure}

Figure~\ref{fig:histo_g0} shows the distribution of extinction-corrected $g$ magnitudes, $g_0$, for all variable stars found in this work. For reference, we show the distribution of magnitudes in the central field from \citetalias{vivas20a} in the top histogram. The RRLS in the central field of CraII have mean $g_0 = 20.91$ with a small dispersion. AC stars in the central field are concentrated around $g_0 \sim 20.0$. The few stars brighter than $g_0=19.0$ are field RRLS along the line of sight. The bottom histogram in Figure~\ref{fig:histo_g0} shows the distribution of magnitudes of the variable stars found in this work. The histogram shows an important number of RRLS at the expected magnitude for CraII confirming the galaxy extends beyond $2\,r_h$ (the region studied in \citetalias{vivas20a}). It is evident, however, that the dispersion of magnitudes of stars associated with CraII (the group around $g_0\sim 21.0$) is much wider than in the central field, likely indicating a dispersion in distances. The bright tail of this group of stars should contain AC stars, which we study in more detail in \S~\ref{sec:AC}. Understandably, the number of foreground field stars (stars with $g_0<19.0$) is much larger here than in \citetalias{vivas20a} since we surveyed an area 13 times larger. Based on this histogram we tentatively selected all 46 stars with $g>19.0$ as potential members of CraII.

\begin{figure}
\includegraphics[width=0.49\textwidth]{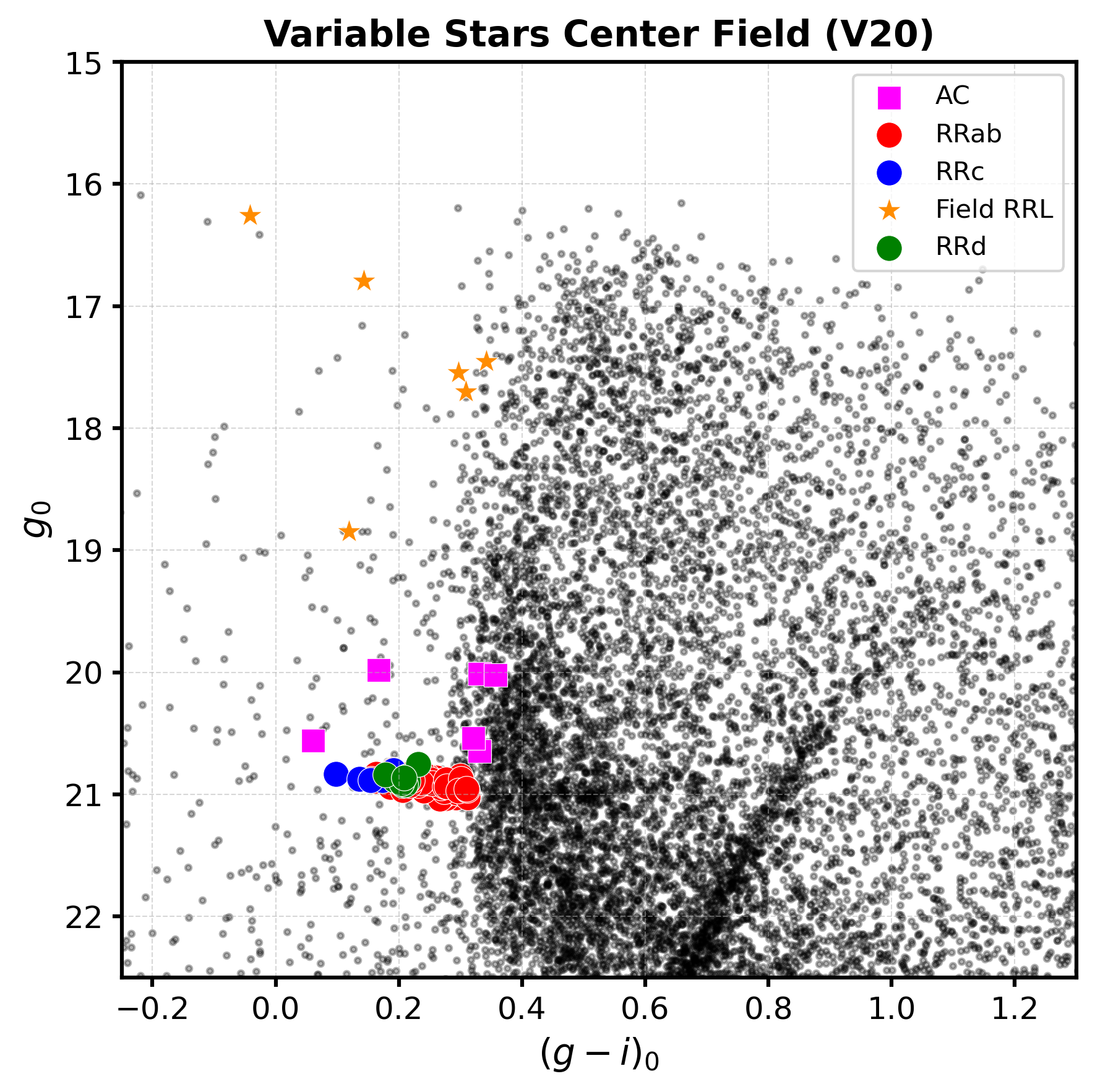}
\includegraphics[width=0.49\textwidth]{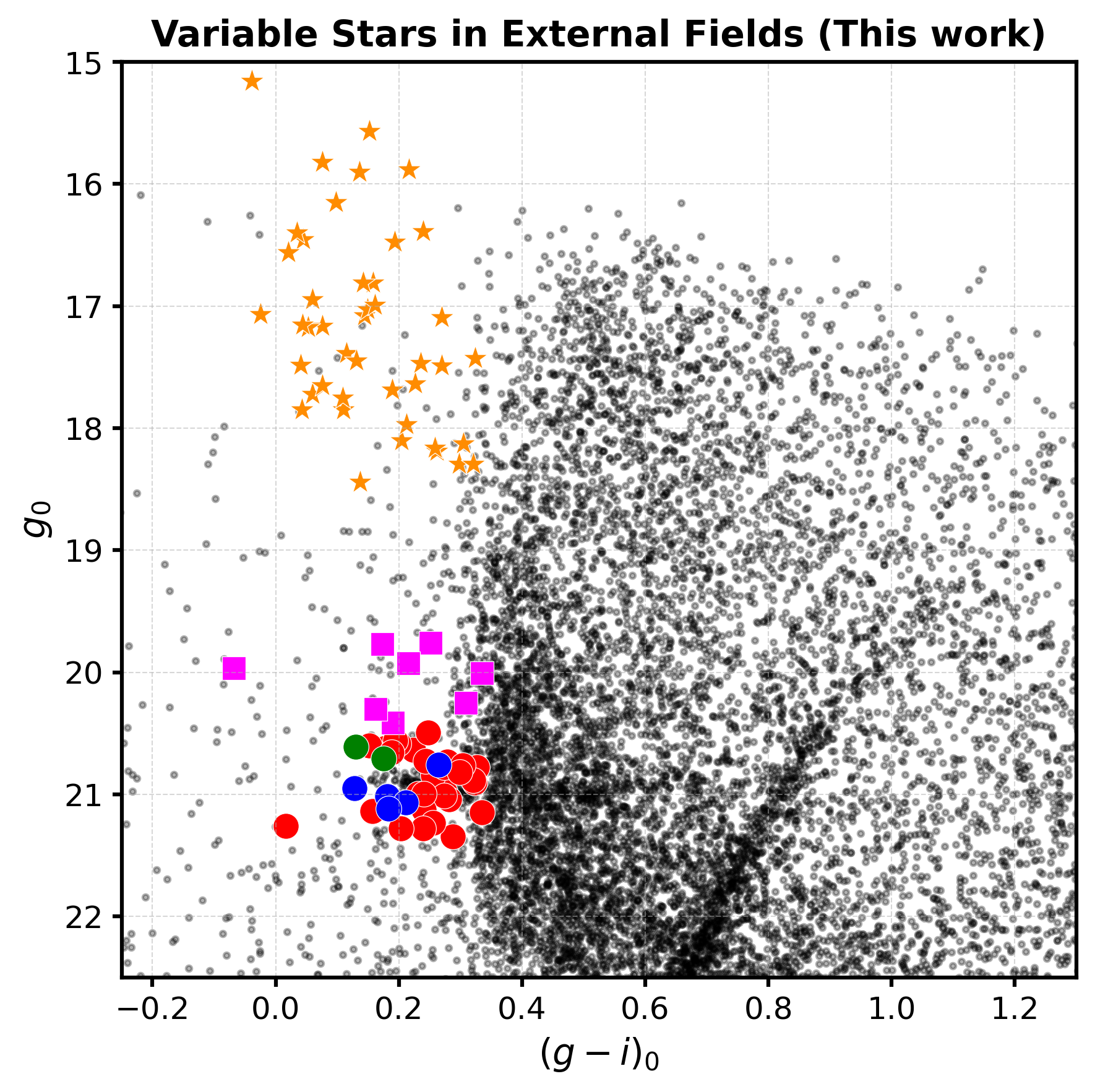}
\caption{Location of the RRLS and AC in the CMD of CraII. The left panel shows with colored symbols the variable stars in the central field observed by \citetalias{vivas20a}, and the right panel contains the stars identified in the external fields studied in this work. The background is shown as a reference of the CraII CMD and it is the same in both panels. It contains all stars within $1\degr$ from the center of CraII, as measured by \citet{walker19}.}
\label{fig:cmd}
\end{figure}

Figure~\ref{fig:cmd} shows the upper part of the CMD of CraII \citep[from][]{walker19} with the RRLS and AC stars overplotted as colored symbols. Here we can see again that the outskirt fields (right panel) have plenty of RRLS in the region of the horizontal branch but display a larger dispersion in magnitude compared with that for the RRLS in the central region (left panel). The background CMD is the same in both panels; it serves as a reference for the main CMD features of the central part of CraII. These features (red giant branch and horizontal branch) cannot be identified in the outskirt fields since the density of CraII stars is low compared with the fore/background.

The extended population of CraII RRLS is composed of 31 \rrab, 5 \rrc, and 2 {\rrd } stars. Of those, 24 RRLS had already been reported by \citet{coppi24}, and thus we are adding here 14 new RRLS in the outskirts of CraII. Our periods agree very well with those reported in \citet{coppi24} for the stars in common, except by two stars. For V166 we report a period of 0.29883~d while \citeauthor{coppi24} found a period of 0.42631~d. The second discrepancy is for star V188 which we classify as a {\rrab } star with period 0.72474~d, while \citeauthor{coppi24} classified it as a {\rrc } star with period 0.42135~d. V188 looks indeed noisy in our data which may be a signal of possible double pulsation mode but more data would be needed to confirm this. Only three of the RRLS in the outskirts of CraII appear in the Gaia DR3 RRL catalog \citep{clementini23}.

Similar to the central field, the outskirts (Figure~\ref{fig:cmd}, right) also contain a majority of {\rrab } stars since the horizontal branch does not extend more to the blue where the hotter {\rrc } stars are produced. The mean period of the {\rrab } stars is 0.65~d, similar to the stars in the central field (0.64~d). 

\subsection{Spatial Distribution} \label{sec:rr_spatial_distr}

\begin{figure}
\plotone{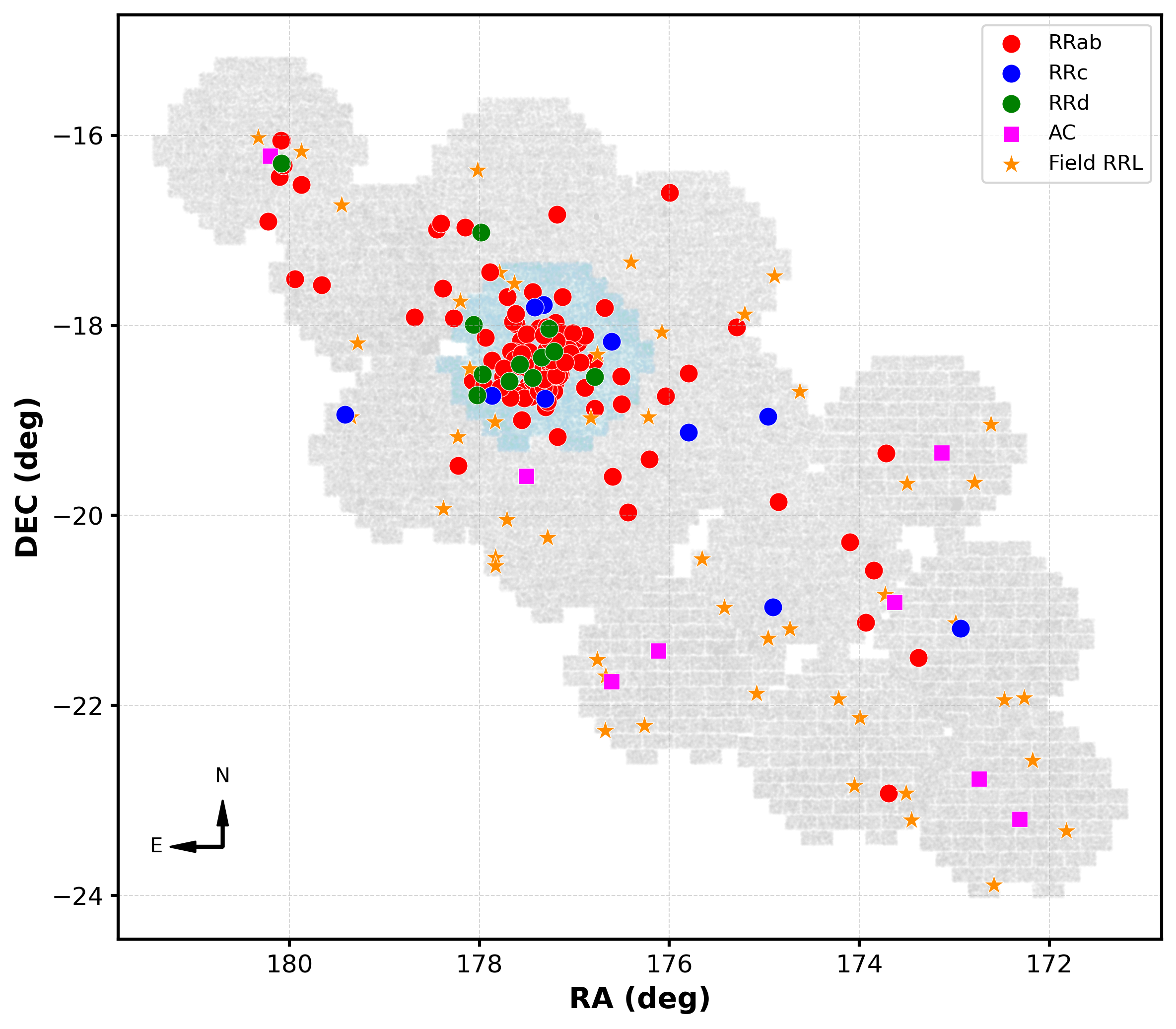}
\caption{Equatorial coordinates of the different type of RRLS and AC stars in CraII \citepalias{vivas20a} and its outskirts (this work). Symbols are the same as in the CMDs of Figure~\ref{fig:cmd}. The gray background is a density map made with all stars in our catalog of outskirts fields. Stars over the blue background are those reported in \citepalias{vivas20a} for the central field.  }
\label{fig:spatial_distr_sky}
\end{figure}

\begin{figure}
\plotone{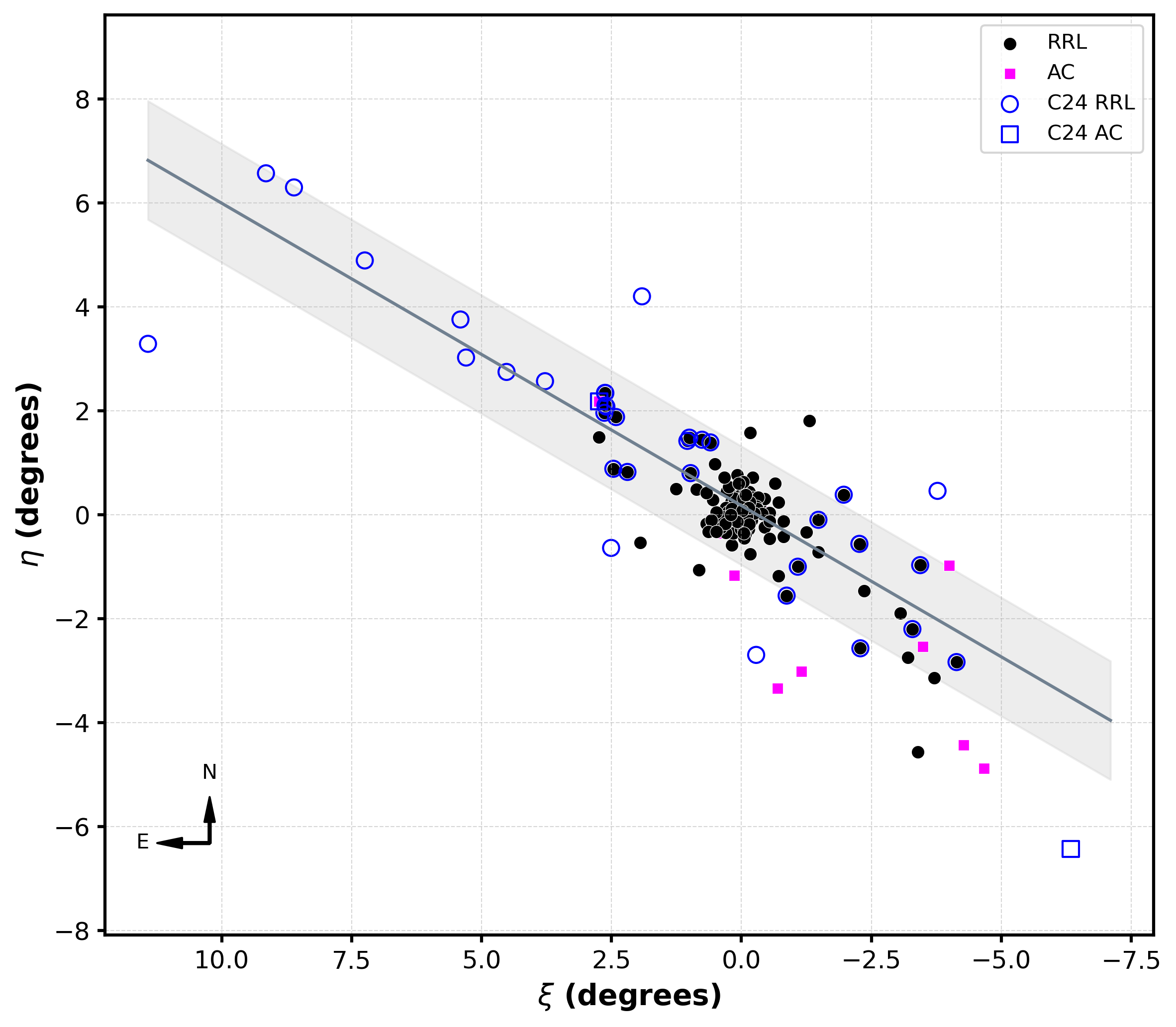}
\caption{Spatial distribution of the RRLS and AC stars in planar coordinates with the center of CraII at $(\xi,\eta)=(0,0)$. Coordinate $\xi$ increases toward the East (to the left in this plot). Stars from \citetalias{vivas20a} and this work are shown as black solid circles (RRLS, all types) and magenta squares (AC). Stars in the tails of CraII identified in La Silla-QUEST survey by \citet[][their Table 2]{coppi24} are shown as open blue circles (RRLS, all types) and blue open squares (AC stars). The solid line and grey region is the best linear regression, and its $1\sigma$ error, of the RRLs in the outskirts (not including the center field).}
\label{fig:spatial_distr}
\end{figure}

Figure~\ref{fig:spatial_distr_sky} shows the spatial distribution of all the variable stars found in CraII from this work and \citetalias{vivas20a}. CraII RRLS were found in all observed fields except in the most external one (ER7). However, there are 2 potential CraII AC stars in that distant field. The most distant RRLS found in our survey is at an angular distance of $5\fdg 7$, or $11\, r_h$ from the center of CraII. It is noticeable that the majority of the RRL stars in the outskirts of CraII have a very elongated distribution, although this may be due in part to the shape of our surveyed area (see Figure 1). The elongated shape is confirmed, however, with the results from La Silla-QUEST survey by \citet{coppi24}. They uncovered RRLS around CraII in an uniform area spanning $170\degr \leq \alpha \leq 190\degr$, $-25\degr \leq \delta \leq -10\degr$, much larger than the region explored by us in this work. In Figure~\ref{fig:spatial_distr} we plot the spatial distribution of all the RRLS and AC stars from this work, from \citetalias{vivas20a} (Center field), and from \citet{coppi24}, in planar coordinates. For clarity, all types of RRLS (\rrab, \rrc, and \rrd) consistent with being members of CraII based on their position in the CMD in \citetalias{vivas20a} and this work are shown as black circles. Similarly, all RRLS (of any type) selected by \citeauthor{coppi24} as members of the tails of CraII are plotted as blue open circles. This plot shows more clearly the long tails in each side of the galaxy, particularly prominent in the North-East side of CraII. We recovered all of the \citeauthor{coppi24}'s stars within our footprint (26 stars). We note, however, that we detected 20 stars (14 RRLS and 6 AC) in the outskirts of CraII which were not found by \citeauthor{coppi24} (black circles or magenta squares with no overlapping open blue symbols)\footnote{For clarity we have not marked with blue open circles the RRLS recovered by \citet{coppi24} in the central field.}. Out of the 20 new stars, 15 are located in the more distant South-West tail. This is understandable given the deeper data of our survey compared with \citeauthor{coppi24}'s.

In total, we found 26 stars in the South-West tail (negative $\xi$'s), and 20 stars in the North-East tail (positive $\xi$'s). The increased number of known stars in the outskirts of CraII allows a better modeling of the tails. 

A linear regression of the position of RRLS in the outskirts of CraII \citep[both from this work and][]{coppi24} determines the direction of the tails. The inclusion of \citeauthor{coppi24}'s stars is important particularly because of the much larger extension they found in the North-East tail. We did not use the AC stars since they are more prone to contamination by field stars (see Section~\ref{sec:contamination}). We obtained a good linear correlation defined by the RRLS,  with Pearson correlation coefficient of 0.85, which is indicated by the solid line in Figure~\ref{fig:spatial_distr}. The position angle of the tails is $59\fdg 8 \pm 1\fdg 9$. Although the stars in the central field were not included in this fit, the model goes very close to the center of CraII (intercept with the y-axis is only $0\fdg18 \pm 0\fdg 14$). Including the stars in the central field of CraII (from \citetalias{vivas20a}) in the linear regression fit produces no significant changes (position angle $= 59\fdg 9 \pm 1\fdg 3$, intercept with the y-axis $= 0\fdg 09 \pm 0\fdg 06$. 

\begin{figure}
\plotone{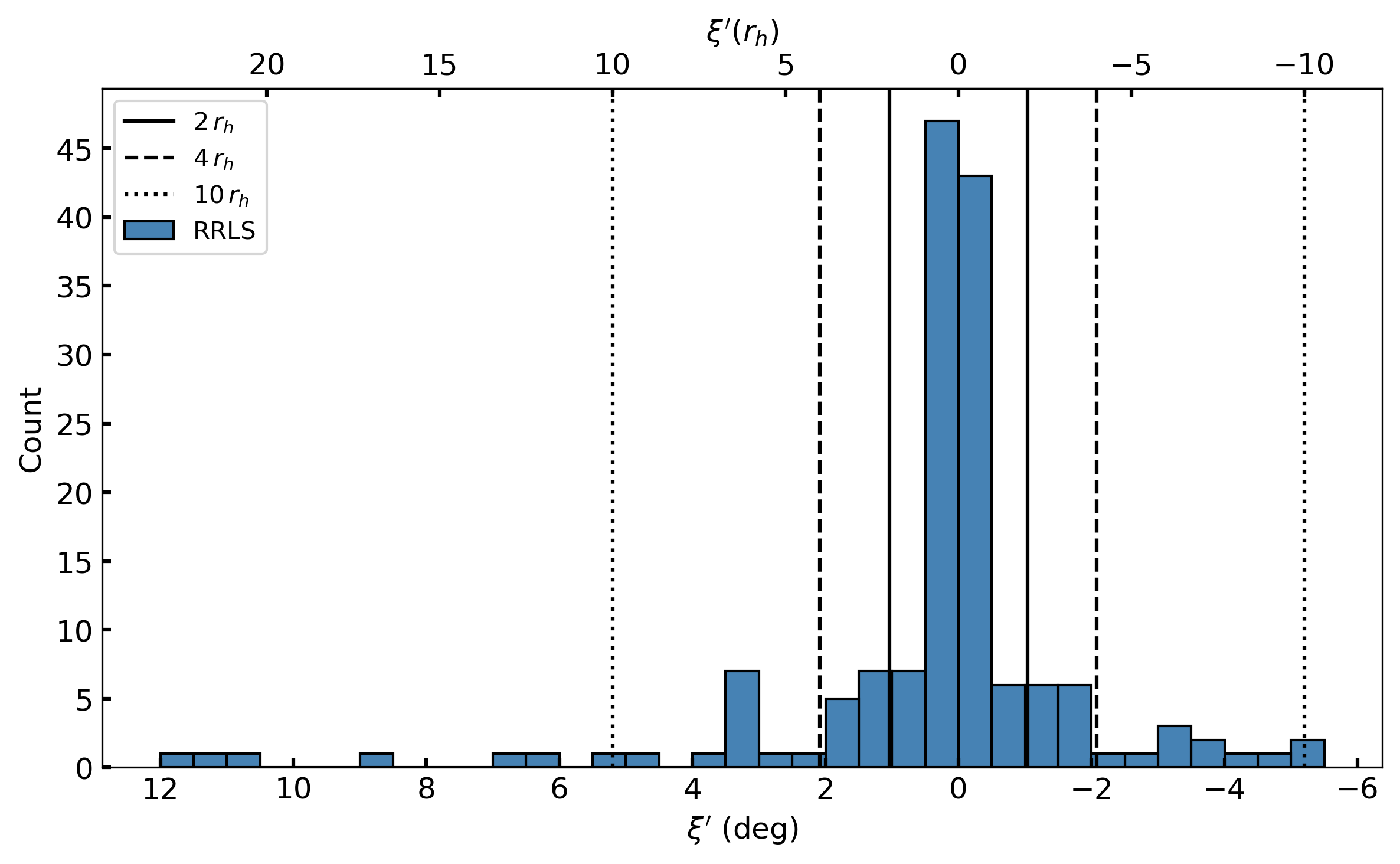}
\caption{Number of RRLS in CraII along the axis of the tails, in bins of $0\fdg 5$. The sample of RRLS include those from \citetalias{vivas20a}, from \citet{coppi24}, and from this work. For convenience $\xi'$ is shown in units of degrees (bottom axis) and in units of $r_h$ (top axis)}
\label{fig:density}
\end{figure}

We study the number density of RRLS in the tails in Figure~\ref{fig:density}, in which we show the number of stars along the tails in bins of 0.5~deg. ($\xi'$ is the $\xi$ coordinate rotated by the position angle of the tails defined in Figure~\ref{fig:spatial_distr}). This histogram shows that the RRLS in CraII are centrally concentrated, with most of the RRLS inside $\sim 1.5 \, r_h$. The tails have an approximate uniform number density of $\sim 1-2$ RRLS/deg outside $4\, r_h$. A noticeable exception occurs at $\xi'=3.25$ deg, where a high density of RRLS is present. This overdensity of stars is made of 7 RRLS located in field ER2 (see Figures~\ref{fig:fields} and \ref{fig:spatial_distr}), concentrated in a region of 50\arcmin of diameter (indeed 6 out of the 7 stars are concentrated within a 30\arcmin region). The central coordinates of the clump are $(\alpha,\delta)=(180\fdg 073,-16\fdg368)$. The RRLS in this clump have an average distance of 103~kpc with a dispersion of 4~kpc. Density variations along streams of disrupted stellar systems are known to exist and may have different causes including the properties of the orbit, the potential of the Milky Way, interactions with low-mass sub-halos, among others \citep{bonaca25}. With the current data we cannot exclude this overdensity is part of an independent, undetected UFD galaxy. Follow-up on this group is encouraged. 

\subsection{Distances and Distance Gradient} \label{sec:gradient}

In \citetalias{vivas20a} we showed that the RRLS in the center of CraII follow a tight Period-Luminosity relationship in the $i$ band but such a tight relationship was not observed in the $g$ band. There was also a larger dispersion in the magnitude distribution of the RRLS in $g$ compared with $i$. Similar behavior is also expected from theoretical relationships \citep{caceres08,marconi22}. Thus, following \citetalias{vivas20a}, we calculate distances to individual RRLS using Period-Luminosity-Metallicity (PLZ) relationship of \citet{caceres08} only in the $i$ band. We also note that the i-band is less sensitive to systematic errors in the mean magnitude due to uncertain amplitude measurements from sparse lightcurves because amplitudes in this band are always smaller than in the bluer bands. For the CraII stars we assumed a metallicity of [Fe/H] $= -2.0$, and an $\alpha$ enhancement of $+0.3$ dex, while for field RRLS, we assumed [Fe/H] $= -1.65$, and an $\alpha$ enhancement of $+0.2$ dex, which are appropriate values for the halo \citep[eg.][]{suntzeff91,pritzl05}. Individual distances and their errors are reported in Table~\ref{tab:data}. 

\begin{figure}
\includegraphics[width=0.49\textwidth]{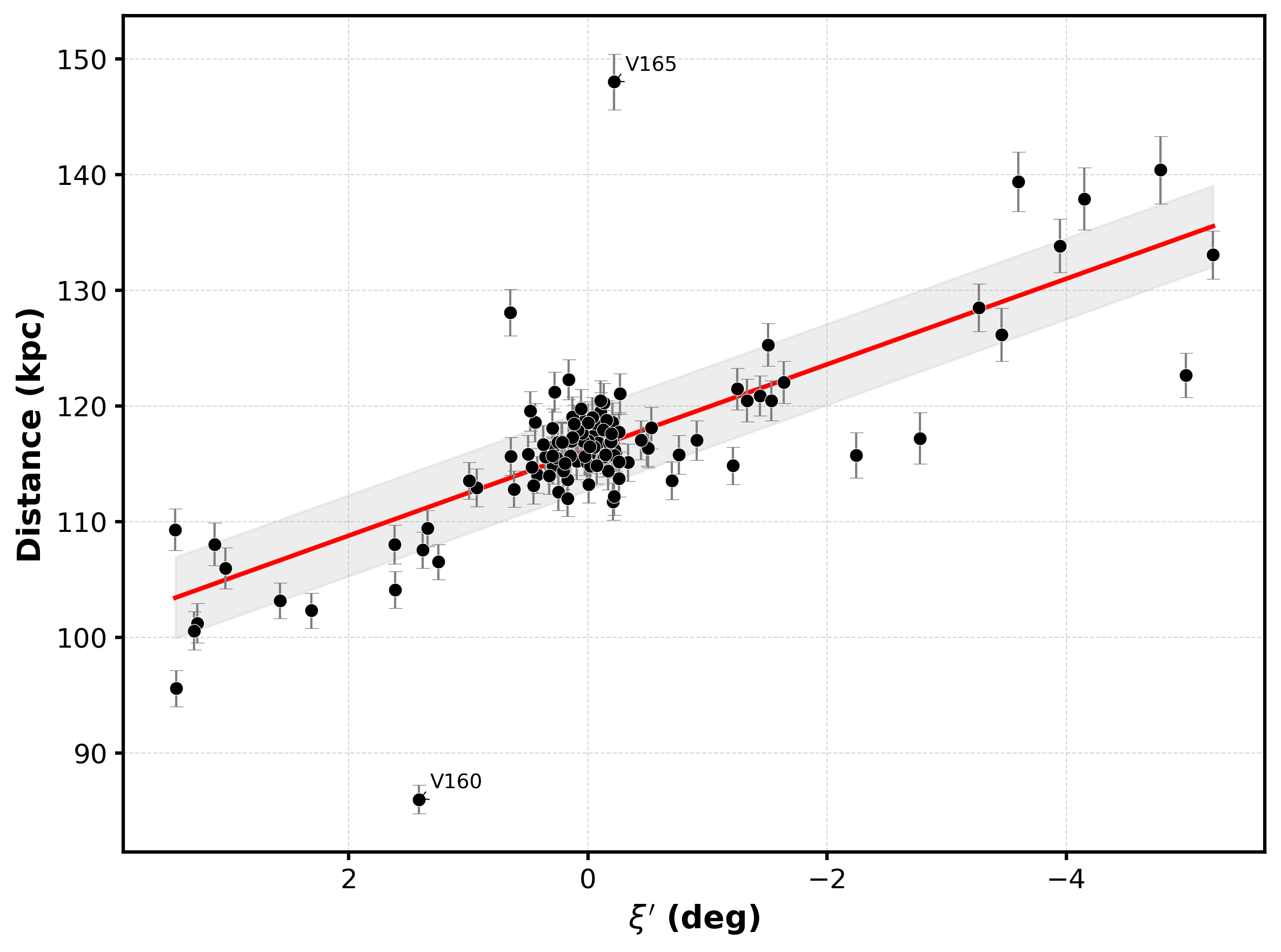}
\includegraphics[width=0.49\textwidth]{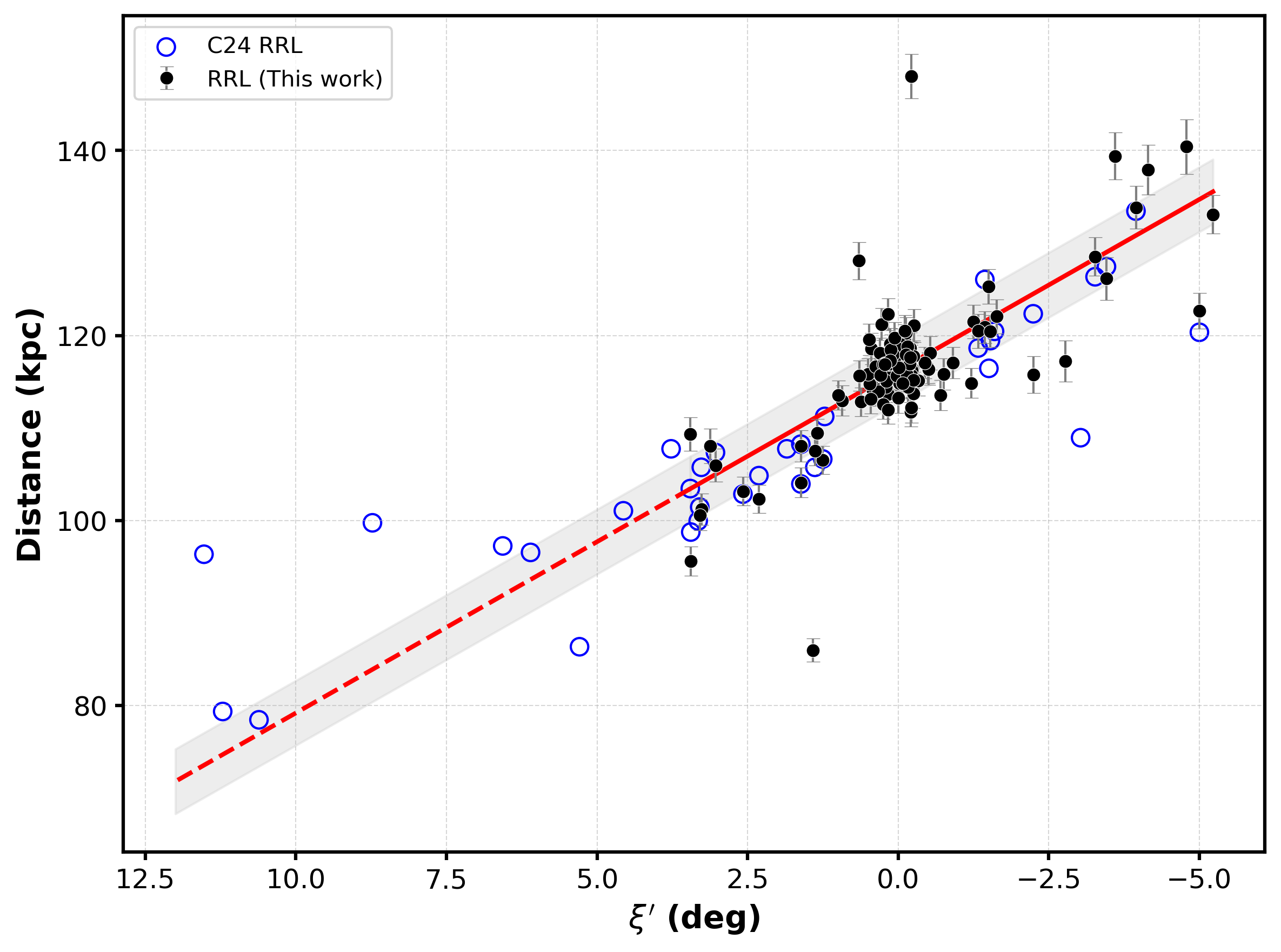}
\caption{(Left) Distance to the RRLs along the stream as defined by the line in Figure~\ref{fig:spatial_distr} (right). The red line is the best fit to the data excluding the likely outlier at $(\xi',d)=(1\fdg 4, \, 86\, {\rm kpc})$ (see text). (Right) Same as left plot but expanding toward larger angular distances in the "near" tail to include the stars detected by \citet{coppi24} (an offset of $-2.78$ kpc was applied to those stars to match our distance scale). The dashed red portion of the line is the extrapolation of our fitted gradient toward larger $\xi'$. }
\label{fig:distance}
\end{figure}

In Figure~\ref{fig:distance} we show the distance to the CraII RRLS (only types ab and c) found in this work along $\xi'$, the rotated $\xi$ coordinate. We excluded the {\rrd } stars since they do not follow the same Period-Luminosity-Metallicity (see Figure 8 in \citetalias{vivas20a}). The distance gradient is clear and confirms the findings by \citet{coppi24}: the North-East tail ($\xi' > 0$) is significantly closer to us than the tail on the other side of the galaxy. The RRLS in the outskirts of CraII have distances from $\sim 95$ to $\sim 140$~kpc, a range of 45 kpc difference along $8\fdg 7$ in the sky. That difference in distance is very much larger that the typical error bar for the distances of individual stars.

The observed distance ($d$) gradient is well modeled by a simple linear fit, $d = (-3.85 \pm 0.31)\, \xi' + (116.2\pm 0.45)$. The rms of the fit is 5.0 kpc. 

There are only two stars in our sample that seem to lie $3\sigma$ away from the trend in distance of the tails. Star V160, in field R6, has a distance of 86~kpc. V160 is a {\rrc} star with a period of 0.16~d. Such a short period is not observed among the {\rrc} stars in the central field \citepalias{vivas20a}, where those stars have a mean period of 0.41~d with a small dispersion of 0.02~d. The very short period, which is rare for halo RRLS, suggest the period may be incorrect (for example, we may have recovered an alias of the right period) or a misclassification. Eclipsing binaries or SX Phe stars are other types of variable stars that can have such short periods. The other suspect star is V165, the most distant star in our sample, located in field R3, at 148~kpc from the Sun, and $2\fdg 2$ from the center of CraII in approximately a direction perpendicular to the tails. This is a {\rrab } star with a period of 0.558~d, which is within the expected range for CraII stars. Although it is very distant, the lightcurve is of high quality (Figure~\ref{fig:lc}). At this very large distance, contamination by field stars is unlikely but not impossible (see next section). Thus, although V165 does not follow exactly the distance gradient trend, we do not discard the possibility that it could be part of the extended population of CraII. 

Discarding both V160 and V165 and recalculating the linear fit we obtain that in the range $-5.3 < \xi' < 3.5$ the distance gradient is defined by

\begin{equation}
 d (\rm {kpc}) = (-3.70\pm 0.21)\, \xi' + (116.2 \pm 0.32)   
\end{equation}

\noindent
with a rms of 3.5~kpc. Repeating the fit but using only stars in the tail (i.e. excluding the \citetalias{vivas20a} stars) results in virtually the same result ($d = (-3.86 \pm 0.35)\, \xi' + (115.2 \pm 0.96)$; rms$=5.5$~kpc). This indicates that, contrary to what was found by \citet{coppi24}, we find no offset in distance between the tails and the center of the galaxy. \citet{coppi24} found an offset of $\sim 5$ kpc between the core and the tails which they attributed to possible population differences. In our data the fitted gradient goes right through the middle of the distribution of the central stars, and thus we find no clear evidence for such population differences here. The reason for this discrepancy between the two studies may be due in part to the higher completeness of the far-side tail in our deeper data.

The distance gradient along the tails expressed in terms of distance modulus is

\begin{equation}
 \mu_o (\rm {mag}) = (-0.072\pm 0.006)\, \xi' + (20.296 \pm 0.018);  \;\;\; \sigma = 0.10  
 \label{eq:mu}
\end{equation}

In the right panel of Figure~\ref{fig:distance} we extrapolated the observed gradient toward the larger (positive) values of $\xi'$, to include the much longer tail detected by \citet{coppi24} in that direction. The extrapolation of the gradient is indicated by the dashed red line. The distance scale between \citet{coppi24} and our work is similar but not identical. Using the 24 RRLS stars in common, we estimated an offset of 2.7~kpc between both works, with our distances being shorter than \citeauthor{coppi24}'s. The discrepancy is likely due to the use of different Period-Luminosity-Metallicity relationships. We applied that offset to \citeauthor{coppi24}'s stars when plotting them in Figure~\ref{fig:distance} (right). The figure shows that although the extended "near" tail follows the trend of the gradient, the blue circles with $\xi' \gtrsim 7\degr$ lie above the red line. Thus, there is an indication that the distance gradient may not be as steep at large distances from the center of CraII as in the inner 5\degr.

In appendix~\ref{sec:appendix} we derive the mean distance to the center of CraII and the distance gradient using different PLZ relationships for the RRLS. The distance scale assumed here using \citet{caceres08} tends to be on the short side compared with others, but the slope of the distance gradient does not significantly change when using other PLZ relationships. 

\subsection{Expected Contamination} \label{sec:contamination}

Field halo RRLS exist at very large distances from the Galactic center \citep{medina18,stringer21,medina24,feng24} but their number density decreases with galactocentric distance with a steep power law. Thus, they become rare in the outer halo. For example, \citet{medina24} found only 23 RRLs with distances $d> 100$~kpc in an area of 350~sq. deg. 

We use the power law determined in \citet{medina24} to model the number density radial profile of RRLS until 200~kpc, which has a slope $n=-4.47$, to estimate how many of the RRLS may be random field stars along the line of sight of our footprint. Our effective observed area in the sky spans 35.7~sq. degrees. In that area, we estimate there should be 1.2 RRLS between 90 and 140~kpc, the range of distances seen in the tails of CraII. Since we found 43 RRLS in that range of distance, the contamination by halo stars is expected to be very small ($\sim 3\%$), and thus will not affect the results presented in the previous sections.

At larger distances, the predicted number of field halo stars is even lower. Between 140 and 160~kpc, we expect only 0.3 RRLS in the observed area. Thus, this makes unlikely (although not impossible) that star V165 discussed in the previous section is a random halo star. Without additional information like proper motions, radial velocities and/or metallicities, it is not possible to confirm the nature of V165. 

We also estimate the number of expected RRLS in the range 65-95~kpc. This would be the range of distances for RRLS within the magnitude range $19.7 \lesssim g_0 \lesssim 20.3$ which correspond to the variables we selected as AC stars. Some of those stars may indeed be field RRLS in the front of CraII. Indeed, from the integration of the power law, we expect 3.3 ($\pm 1.8$) RRLS in our survey area, meaning that the contamination of the much smaller AC stars sample by field stars may be as large as $3.3/8$, or 41\%. In the next section we discussed the membership of the AC stars with more detail.  

\section{Anomalous Cepheids} \label{sec:AC}

\begin{figure}
\plotone{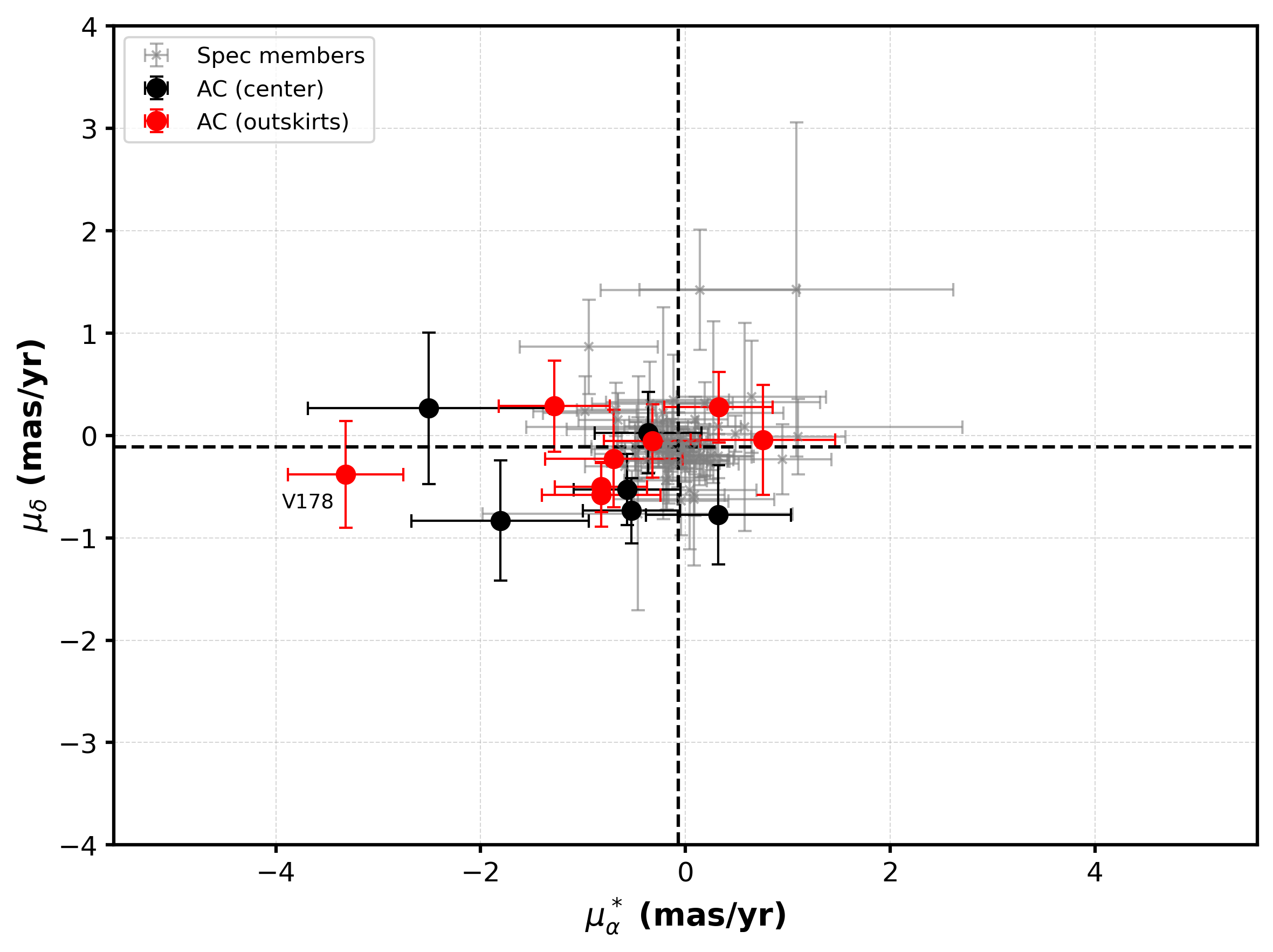}
\caption{Proper motion of radial velocity members of CraII (grey asterisks) and AC stars in the center field (black circles, from \citetalias{vivas20a}), and in the outskirts (red circles, this work). The spectroscopic members were taken from \citet{caldwell17} and \citet{fu19}. Dotted lines indicate the CraII proper motion from \citet{battaglia22}.}
\label{fig:ACpm}
\end{figure}

The pulsation properties (light curves, amplitudes, periods) of AC stars have a large overlap with those of RRLS. Therefore, it is very hard to separate AC stars in the field from the more abundant RRLS. In stellar systems, the position of the stars in the CMD, up to $\sim 2$ mag above the horizontal branch is an indication that AC stars are present. The analysis of the variable stars in the central field \citep[\citetalias{vivas20a},][]{joo18,monelli18,coppi24} of CraII indicates the galaxy has 6 AC stars\footnote{In \citetalias{vivas20a} we suggested V108 was an AC star. However further analysis by \citet{ngeow22AC} and \citet{coppi24} rejected it as being an AC star. Consequently, in this work we change the classification of V108 to \rrab.} (Figure~\ref{fig:cmd}, left). Although some AC stars would be expected in the tails, it is puzzling to find 8 of them, more than the number in the center. It would be expected that the ratio between the number of RRLS and AC stars is approximately the same. Furthermore, two of those stars are farther away than RRLS in one of the tails, and we note that \citet{coppi24} found another AC in the same tail even farther away, beyond our coverage (see Figure~\ref{fig:spatial_distr}.) The correct identification of AC stars in the tails of CraII has additional challenges. First, contamination by field RRLS is possible (see Section~\ref{sec:contamination}), and second, the distance gradient of the tails makes it harder to establish the location of the AC candidates with respect to the horizontal branch of CraII at that location. In this section we tried several ways to address the membership of the AC candidates identified in the tails and shown in the CMD of Figure~\ref{fig:cmd} (right panel).

The AC stars are bright enough to have good proper motion measurements from Gaia DR3 \citep{gaiadr3}. Figure~\ref{fig:ACpm} shows that all our candidate AC stars, except V178, have proper motions consistent with the spectroscopic members (mostly red giant stars) in the center of CraII \citep[from][]{caldwell17,fu19}. V178 is more than $5\sigma$ away from the mean proper motions of CraII, $(\mu_\alpha^*,\mu_\delta) = (-0.07\pm0.02, -0.11\pm0.01)$ mas/yr, as determined by \citet{battaglia22}.

\begin{figure}
\plotone{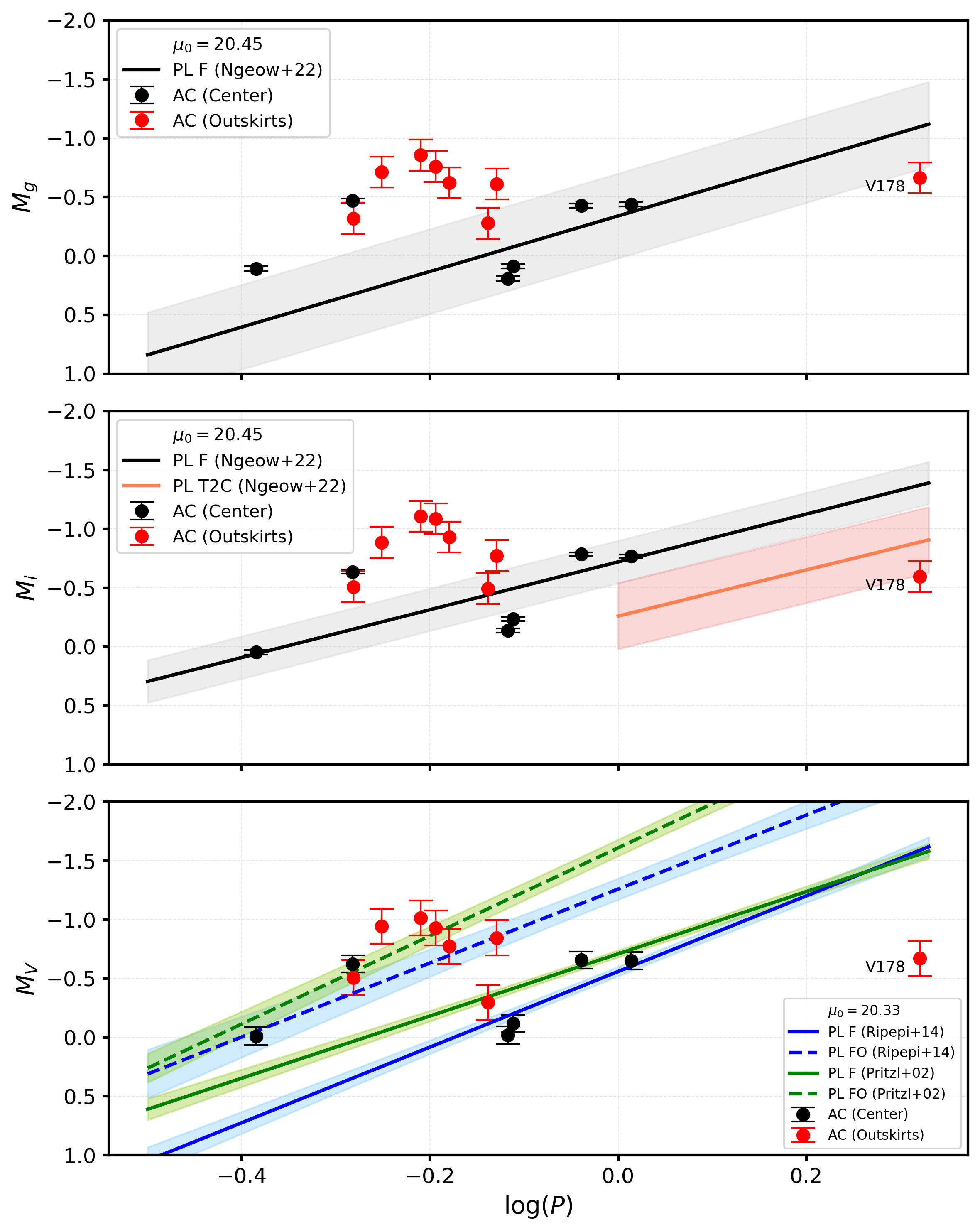}
\caption{Absolute Magnitude versus period for the AC stars in the central field of CraII (Black circles, from \citetalias{vivas20a}), and in the tails (red circles) in 3 different bands, $g$ (Top), $i$ (Middle) and $V$ (Bottom). For $g$ and $i$ we overlay the PL relationships for Fundamental-mode AC stars from \citet[][]{ngeow22AC} as black solid lines. In the bottom panel, two different PL relationships in the $V$ band for Fundamental (solid lines) and First Overtone (dashed lines) AC stars are shown \citep{pritzl02,ripepi14}.}
\label{fig:AC_PL}
\end{figure}

The AC stars obey Period-Luminosity (PL) relationships. We can use this property to investigate if our AC candidates are indeed located at the distance of CraII or their tails. By anchoring the AC stars to the distance of the RRLS, we can investigate if they obey the expected PL relationship. 

\citet{ngeow22AC} defined PL relationships using  Zwicky Transient Facility \citep[ZTF,][]{ztf} data which with very small offsets (only in the $g$ band) are compatible with the SDSS system to which our data is tied to \citep[see details in][]{ngeow22AC}. Unfortunately, only PL relationships for Fundamental-mode (F) pulsators are known in these bands. In the top and middle panels of Figure~\ref{fig:AC_PL} we show the absolute magnitudes $M_g$ and $M_i$ as a function of the logarithm of the pulsational period for the AC in the central field (from \citetalias{vivas20a}, black circles) and in the tails of CraII (red circles). For the latter, we used the $\xi'$ position of the AC stars to predict the distance at that position in the tail using Eq.~\ref{eq:mu}, with a small offset to be compatible with \citet{ngeow22AC}. In their work, they found a slightly larger average distance modulus ($\mu_0 = 20.45$) when applying their PL to the CraII in \citetalias{vivas20a}, compared with the distance given by the RRLS in \citetalias{vivas20a} ($\mu_0 = 20.33$). We used \citeauthor{ngeow22AC}'s distance in Figure~\ref{fig:AC_PL} (top and middle panels) to reproduce the location of the AC stars in the central field, where they found that the three AC star with the largest period fitted very well the F-mode relationship. We indeed reproduced that result, except that now star V107, for which we revised the period to 1.0318~d following the results by \citet{coppi24}, now also fits in the F-mode line of the $g$ band\footnote{Our initial period search in \citetalias{vivas20a} had indeed suggested the period of 1.0318~d, which we adopt here, as one that produced a similarly good light curve as our original period of 0.5134~d.}. One of the stars in the tails, V186, fits within 1$\sigma$ the F-mode PL relationship. Most of the other stars seem to be too bright for that PL relationship, although they may be pulsating in the FO mode.

To explore this possibility we transform our $g,i$ magnitudes to Johnson-V, using the transformation equations provided in \citet[][from SDSS to DES, and then from DES to Johnson V]{desDR2}. The transformation equations errors were budgeted in the final $M_V$ calculation. Extinction was calculated using $A_V = - 2.742\, \rm{E(B-V)_{SFD}}$ \citep{schlafly11}. The bottom panel shows the transformed data together with 2 different PL relationships in this band, \citet{pritzl02} in green, and \citet{ripepi14}, which fit very well our data using our RRLS distance modulus of $\mu_0 = 20.33$. Indeed, the brightest stars are compatible with being pulsating in the FO mode. Although there are some small differences between the two PL relationships, we can say that all but one of the AC stars in the tails are compatible with being members of CraII based on their agreement with the PL relations anchored to the distance of the RRLS. Surprisingly, it seems that only one star in the tails is pulsating in the F mode, which is expected to be the most common mode \citep[for example, out of 250 AC stars in the Magellanic Clouds, 70\% pulsate in the F mode,][]{soszynski15}. On the other hand, out of the 11 AC stars found in Leo~T, 8 are found to be pulsating in the FO mode \citep{clementini12,ripepi14}, and thus the apparent distribution of modes in the CraII tails may not be completely unexpected. 

Only one star does not seem compatible with the PL relationships for the AC stars at the distance of CraII. This star is the one with the largest period; it appears to be too faint for that period, particularly in the $i$ and in the V band. No surprise, this star is V178, which is also a proper motion outlier. With a 2.1~d period, V178 does not seem to be a field RRLS since this period is much too long for that type of star, although we note that the phase coverage of the observations is not great and a period of 1.107~d produces a reasonably light curve as well. V178 also has a much bluer color than the rest of the AC stars (see Figure~\ref{fig:cmd}, right), and it definitely would be a color too red to be a {\rrab } star. If the 2.1~d period for V178 is confirmed, it could be a field Type II Cepheid of the type BL Her. Unfortunately this star was not recovered in \citet{coppi24}; it appears in the ZTF DR23 database{\footnote{\url https://irsa.ipac.caltech.edu/Missions/ztf.html}, under IDs 269215300000440 and 269115300002623 in the $r$ and $g$ band, respectively. The star shows variability but not periodicity in those data. With the current information we are unable to say more about the nature of this star. 

\section{Discussion and Conclusions} \label{sec:conclusions}

We searched for RRLS and AC stars in a region of $35.7$ sq deg around the CraII dwarf galaxy with DECam.
Out of the 46 periodic variable stars with $g_0>19.0$, 44 (37 RRLs, 7 AC) seem to be genuine members of the CraII galaxy. About 60\%, or 26 of those 44 variables (20 RRLS, 6 AC) are located at more than $4\, r_h$ from the center of CraII confirming the presence of a significant extended stellar population in CraII. Although we covered a smaller area than the La Silla-QUEST survey, our improved depth compared to that survey has allowed us to unveil more variable stars in the region and define better the tidal tails of CraII. Our stars span almost $10\degr$ in the sky across the galaxy. The RRLS show a strong gradient in distance of 3.7~kpc/deg for the extratidal material with the South-West tail reaching distances as far as 140~kpc from the Sun and the North-East tail having stars as close as 80~kpc. 

In the far-side tail, or trailing tail, we found 24 stars (6 AC, 18 RRL), 13 of which are new discoveries. Although we explored a region until 14$r_h$ in the direction of this tail, the most distant RRLS found is located at $10.9\, r_h$. The depth of our data would have allowed us to find more distant RRLS, if they exist. The lack of RRLS beyond that point seem to indicate the tail disappears or the density becomes too low to be traced effectively by variable stars. The are, however, 2 AC stars extending beyond the RRL, up to an angular distance of $6\fdg 7$ or 13$r_h$. \citet{coppi24} found another AC in this tail even farther away ($9\degr$ or 17$r_h$), beyond our coverage. Given that contamination by foreground RRLS may affect the AC sample, it would be important to further study those AC stars to confirm whether or not they are part of the stream.

In the near-side tail, or leading tail, we found 19 stars (1 AC, 18 RRLS), of which 4 RRLS are new discoveries. In this tail, we found RRL up to a distance of 6.7$r_h$. This tail, however, has been better traced by \citet{coppi24} who showed that it extends well beyond our area of coverage, with the more distant RRL at an impressive distance of $11\fdg 7$, or 22.5$r_h$. We found an indication that the gradient in distance in the distant part of this tail may not be as steep as it is in the inner $5\deg$. The leading tail also displays a significant over-density of RRLS located $\sim 3.25$ deg from the center of CraII.

Joining our 37 RRLS in the outskirts of CraII with the 21 RRLS found by \citet{coppi24} outside our footprint, there are 58 RRLS outside 2$r_h$ of CraII. Considering that there are 99 RRLS inside 2$r_h$ \citepalias{vivas20a}, we can approximately establish that the outskirts of CraII contain as much as $\sim 60\%$ of the central mass of CraII. 

This work has demonstrated the power of RRLS to trace old stellar populations in the halo. With the deep coverage of the Vera C. Rubin Observatory's LSST over the whole equatorial and southern sky, it will be possible to efficiently discover and trace extended stellar populations of satellite galaxies in the Galactic halo. Such observations will be a critical input for answering open questions about the formation, evolution and dissolution of these systems.

\begin{acknowledgments}
This project used data obtained with the Dark Energy Camera (DECam),
which was constructed by the Dark Energy Survey (DES) collaboration.
Funding for the DES Projects has been provided by 
the U.S. Department of Energy, 
the U.S. National Science Foundation, 
the Ministry of Science and Education of Spain, 
the Science and Technology Facilities Council of the United Kingdom, 
the Higher Education Funding Council for England, 
the National Center for Supercomputing Applications at the University of Illinois at Urbana-Champaign, 
the Kavli Institute of Cosmological Physics at the University of Chicago, 
the Center for Cosmology and Astro-Particle Physics at the Ohio State University, 
the Mitchell Institute for Fundamental Physics and Astronomy at Texas A\&M University, 
Financiadora de Estudos e Projetos, Funda{\c c}{\~a}o Carlos Chagas Filho de Amparo {\`a} Pesquisa do Estado do Rio de Janeiro, 
Conselho Nacional de Desenvolvimento Cient{\'i}fico e Tecnol{\'o}gico and the Minist{\'e}rio da Ci{\^e}ncia, Tecnologia e Inovac{\~a}o, 
the Deutsche Forschungsgemeinschaft, 
and the Collaborating Institutions in the Dark Energy Survey. 
The Collaborating Institutions are 
Argonne National Laboratory, 
the University of California at Santa Cruz, 
the University of Cambridge, 
Centro de Investigaciones En{\'e}rgeticas, Medioambientales y Tecnol{\'o}gicas-Madrid, 
the University of Chicago, 
University College London, 
the DES-Brazil Consortium, 
the University of Edinburgh, 
the Eidgen{\"o}ssische Technische Hoch\-schule (ETH) Z{\"u}rich, 
Fermi National Accelerator Laboratory, 
the University of Illinois at Urbana-Champaign, 
the Institut de Ci{\`e}ncies de l'Espai (IEEC/CSIC), 
the Institut de F{\'i}sica d'Altes Energies, 
Lawrence Berkeley National Laboratory, 
the Ludwig-Maximilians Universit{\"a}t M{\"u}nchen and the associated Excellence Cluster Universe, 
the University of Michigan, 
NSF’s NOIRLab, 
the University of Nottingham, 
the Ohio State University, 
the OzDES Membership Consortium
the University of Pennsylvania, 
the University of Portsmouth, 
SLAC National Accelerator Laboratory, 
Stanford University, 
the University of Sussex, 
and Texas A\&M University.

Based on observations at Cerro Tololo Inter-American Observatory, a program of NSF NOIRLab (NOIRLab Prop. ID 2020A-0058 and 2021A-0124; PI: K. Vivas), which is managed by the Association of Universities for Research in Astronomy (AURA) under a cooperative agreement with the U.S. National Science Foundation. Part of this research was conducted during the “Research Experience in Optical-Infrared Astronomy at NOIRLab in Chile” REU Site funded by NSF Award Number 2349023. M.Mo. acknowledges support from Spanish Ministry of Science, Innovation and Universities (MICIU) through the Spanish State Research Agency under the grants "RR Lyrae stars, a lighthouse to distant galaxies and early galaxy evolution" and the European Regional Development Fun (ERDF) with reference PID2021-127042OB-I00.
M.Mo acknowledges the INAF projects "Participation in LSST – Large Synoptic Survey Telescope" (LSST inkind contribution ITA-INA-S22, PI: G. Fiorentino), OB.FU. 1.05.03.06 and "MINI-GRANTS (2023) DI RSN2" (PI: G. Fiorentino), OB.FU. 1.05.23.04.02.
CG and M.Mo acknowledge support from the Agencia Estatal de Investigaci\'on del Ministerio de Ciencia e Innovaci\'on (AEI-MCINN) under grant “At the forefront of Galactic Archaeology: evolution of the luminous and dark matter components of the Milky Way and Local Group dwarf galaxies in the {\it Gaia} era” with reference PID2023-150319NB-C21/10.13039/501100011033. D.L.N. acknowledges support from NSF grants AST 1908331, 2109196, and 2408159. We thank Douglas Tucker for providing the transformation equations to LSST bands in advance of their publication. We thank constructive comments from the anonymous referee which helped to improve several parts of this work. 
\end{acknowledgments}

\vspace{5mm}
\facilities{Blanco (DECam)}

\software{
TOPCAT \citep{taylor05}; 
PHOTRED \citep{nidever17};
{\texttt{matplotlib}}\,\citep{matplotlib},}
{\texttt{astropy}}\,\citep{astropy2013,astropy2018};
{\texttt{numpy}}\,\citep{numpy}
}

\appendix

\section{Distances using different PLZ relationships} \label{sec:appendix}

In order to have an estimation of the external uncertainties due to our choice of PLZ, we re-calculated the mean distance to CraII (using the RRLS in the central field from \citetalias{vivas20a}), and the distance gradient along the tails (using the RRLS from this work) using three recent PLZ relationships derived by \citet{ngeow22RR}, \citet{narloch24}, and \citet{marconi22}.

\citet{ngeow22RR} obtained PLZ relationships for {\rrab } and {\rrc } stars in the Pan-STARRS1 photometric system from a large number of RRLS in globular clusters from ZTF, spanning a large range of metallicities. Following \citet{ngeow22RR}, an offset of $-0.003$ and $-0.004$ mag was applied to the PLZ in $i$, for {\rrab } and {\rrc } stars respectively, to transform to the SDSS system in which our CraII data is tied. \citet{narloch24} also obtained PLZ relationships in the Pan-STARRS1 system from 35 nearby field RRLS with measured high-resolution spectroscopic metallicities spanning the range from $-0.03$ to $-2.59$ dex. For the {\rrc } stars we used their combined RRab+RRc relationship since \citet{narloch24} have too few {\rrc } to fit that relationship independently. We used the version of the PLZ based on geometric Gaia DR3 parallaxes. We applied the same offsets \citep[from][]{ngeow22RR} to bring the PLZ relationships to the SDSS system.
On the other hand, \citet{marconi22} used nonlinear convective pulsation models for RRLS, covering a broad range of metal content, to derive PLZ relationships for both fundamental and first overtone pulsators in the LSST filters. In this case we transformed our mean magnitudes in $g$ and $i$ into $i_{\rm LSST}$ using the following equation \citep[Douglas L. Tucker, private communication; to appear in Rubin Technical Note RTN-099,][]{RTN-099}:

\begin{equation}
    i_{\rm LSST} = i_{\rm SDSS} + 0.019 - 0.012 \, (g_{\rm SDSS} - i_{\rm SDSS})
\end{equation}

\noindent
which has a rms of 0.012 mag that was added to the error budget. The extinction in this band was obtained using $A_i ({\rm LSST}) = -2.054 E(B-V)_{\rm SFD}$ following recommendations from the Rubin Community Forum\footnote{\url https://community.lsst.org}. For the {\rrc } stars we used the version of the PLZ with a fixed slope from the fundamental mode pulsators \citep[see Table 4 in][]{marconi22}. 

Table~\ref{tab:PLZ} shows the results for the mean distance of CraII and the distance gradient in the tails using these PLZ relationships. For completeness, we include in the table the results using the PLZ in \citet{caceres08} which are discussed in \S~\ref{sec:gradient}. In all cases we assumed a metallicity of [Fe/H] $= -2.0$. We note that \citeauthor{narloch24}'s PLZ gives a distance to CraII significantly larger than the other PLZ relationships, and a steeper distance gradient along the tails.

\begin{deluxetable}{lcccc}
\tablecolumns{5}
\tablewidth{0pc}
\tablecaption{Distance modulus of CraII and Distance modulus gradient of the tails using different  PLZ relationships\label{tab:PLZ}}
\tablehead{
\colhead{PLZ} & \colhead{$\mu_0$ (CraII)} & \colhead{$\sigma \mu_0$ (CraII)} & \colhead{Distance modulus gradient} & \colhead{rms} \\
\colhead{} & \colhead{(mag)} & \colhead{(mag)} & \colhead{(mag)} & \colhead{(mag)} \\
}
\startdata
\citet{caceres08} & 20.332 & 0.039 & $(-0.072\pm 0.006)\, \xi' + (20.296 \pm 0.018)$ & 0.10 \\
\citet{ngeow22RR} & 20.427 & 0.040 & $(-0.071\pm 0.007)\, \xi' + (20.393 \pm 0.018)$ & 0.10 \\
\citet{marconi22} & 20.464 & 0.041 & $(-0.074 \pm 0.006)\, \xi' + (20.433 \pm 0.017)$ & 0.10 \\
\citet{narloch24} & 20.726 & 0.042 & $(-0.085 \pm 0.007)\, \xi' + (20.729 \pm 0.020)$ & 0.12 \\
\enddata
\end{deluxetable}

\bibliography{CraterII.bib}{}
\bibliographystyle{aasjournalv7}

\end{document}